\documentclass[%
 reprint,
 dvipdfmx,
superscriptaddress,
 amsmath,amssymb,
 aps,
]{revtex4-1}

\usepackage{graphicx}
\usepackage{dcolumn}
\usepackage{bm}
\usepackage{hyperref}

\makeatletter
\newcommand{\colorcaption}[2][]{%
  \begingroup%
  \renewcommand{\@caption@fignum@sep}{ (color online). }%
  \caption[#1]{#2}%
  \endgroup%
}
\makeatother

\begin{document}

\preprint{APS/123-QED}

\title{Seismic Cross-coupling Noise in Torsion Pendulums}

\author{Tomofumi Shimoda}
\email{shimoda@granite.phys.s.u-tokyo.ac.jp}
\affiliation{Department of physics, The University of Tokyo, Hongo 7-3-1, Tokyo 113-0033, Japan}
\author{Naoki Aritomi}
\affiliation{Department of physics, The University of Tokyo, Hongo 7-3-1, Tokyo 113-0033, Japan}
\author{Ayaka Shoda}
\affiliation{National Astronomical Observatory of Japan, Osawa 2-21-1, Mitaka, Tokyo 181-8588, Japan}
\author{Yuta Michimura}
\affiliation{Department of physics, The University of Tokyo, Hongo 7-3-1, Tokyo 113-0033, Japan}
\author{Masaki Ando}
\affiliation{Department of physics, The University of Tokyo, Hongo 7-3-1, Tokyo 113-0033, Japan}
\date{\today}

\begin{abstract}
Detection of low frequency gravitational waves around 0.1 Hz is one of the important targets for future gravitational wave observation.
One of the main sources of the expected signals is gravitational waves from binary intermediate-mass black hole coalescences which is proposed as one of the formation scenarios of supermassive black holes.
By using a torsion pendulum, which can have a resonance frequency of a few millihertz, such signals can be measured on the ground since its rotational motion can act as a free mass down to 0.01 Hz.
However, sensitivity of a realistic torsion pendulum will suffer from torsional displacement noise introduced from translational ground motion in the main frequency band of interest.
Such noise is called seismic cross-coupling noise and there have been little research on it.
In this paper, systematic investigation is performed to identify routes of cross-coupling transfer for standard torsion pendulums.
Based on the results this paper also proposes reduction schemes of cross-coupling noise, and they were demonstrated experimentally in agreement with theory.
This result establishes a basic way to reduce seismic noise in torsion pendulums for the most significant coupling routes.

\end{abstract}

\maketitle

\section{\label{sec:intro}Introduction}
Recent observations of gravitational waves (GWs) and their electromagnetic counterparts have opened a completely new era of astronomy and physics \cite{GW150914,GW151226,GW170104,GW170814,GW170817}.
They enabled us to probe coalescences of binary black holes and neutron stars in an unprecedented way. 
It is essential to enlarge the observational frequency band in the future because GWs from massive black holes are expected to be observed at low frequencies. 
For example, the main frequencies of GWs from merging black holes of $10^3$-$10^6 M_{\rm sun}$, which are called intermediate-mass black holes (IMBHs), are calculated to be around 0.1 Hz.
Observing these mergers will provide important information about the supermassive black hole (SMBH) formation process since IMBH coalescence is one of the possible formation scenarios of SMBHs.

However, the sensitivities of current interferometric GW detectors such as aLIGO \cite{LIGO}, AdVirgo \cite{Virgo} and KAGRA \cite{iKAGRA} are limited to above 10 Hz because they measure GWs through translational motions of suspended test masses.
Translational pendulums typically have resonance frequencies on the order of 1 Hz, therefore the test masses have low response to GWs and also the seismic noise is not filtered by the pendulum below the resonance frequencies. 

For the purpose of observing low frequency GWs, a torsion-bar antenna (TOBA) was recently proposed as a ground-based GW detector using a torsion pendulum \cite{TOBA}.
It measures the rotation of horizontally suspended test mass(es) excited by tidal forces from GWs. 
Since a torsion pendulum can have a low resonant frequency ($\sim$ 1 - 10 mHz), its rotational mode acts as a free-mass at low frequencies down to $\sim$ 0.01 Hz even on the ground, allowing it to respond to low frequency GWs ($\sim$ 0.1 Hz).
Currently planned space-based detectors such as LISA \cite{LISA} and DECIGO \cite{DECIGO} are expected to have extreme sensitivities at low frequencies. Alternatively, TOBA is constructed on the ground with lower cost and better accessibility for upgrading. 
TOBA has a potential sensitivity of $10^{-19} /\sqrt{\rm Hz}$ at 0.1 Hz with 10 m long bars, which is set as the final target.
With this target sensitivity, the observable range reaches 10 Gpc for $10^5 M_{\rm sun}$ IMBH coalescences \cite{TOBA}. 

Additionally, low frequency detectors on the ground can also be used to measure terrestrial gravitational perturbations such as Newtonian noise \cite{NN} and earthquake generated fluctuations.
Newtonian noise from the ground \cite{NNseis} and the atmosphere \cite{NNatomD} is predicted to be significant in next generation GW detectors.
Therefore, direct detection of it is important to understand the nature of the noise and to demonstrate feed-forward subtraction of it \cite{NNcancel}.
Newtonian noise is a promising target even for a smaller ($\sim$ 1 m) TOBA because it is estimated to be $10^{-15} /\sqrt{\rm Hz}$ at 0.1 Hz.
Establishing the mitigation method with a small scale TOBA is also essential to reach the target sensitivity of TOBA. 
Apart from this, detecting gravitational perturbations from transient ground deformation induced by earthquakes can be used for a new early-warning system of earthquakes \cite{Earthquake}, which can be faster than the current method of using seismic P-wave arrival.
Earthquake gravity signals have already been observed during Tohoku-Oki earthquake on 11 March 2011 with gravimeters and seismometers \cite{EEW_JPM2016,EEW_MV2017}.
For the observation of smaller earthquakes, gravity gradient detectors such as TOBA are thought to be necessary in order to filter out seismic noise. 

Several prototypes of TOBA have been developed so far. 
Strain sensitivities of $10^{-8}$ $/\sqrt{\rm Hz}$ at 0.1 Hz \cite{Phase1TOBA} and  $10^{-10}$ $/\sqrt{\rm Hz}$ at 5 Hz \cite{Phase2TOBA} have been achieved with the prototypes.
Though they are successful as a proof of concept for TOBA, their sensitivity is not enough for the scientific observations described above.
One of the largest noises of these prototypes at the main observation band, 0.1 - 10 Hz, was seismic cross-coupling noise, which is displacement noise transferred from translational seismic motion via small asymmetries of the system.
It is essential to reduce this noise sufficiently for more accurate observations.

We have two ways to reduce cross-coupling noise; the first would be suppressing the translational seismic motion with some vibration isolation techniques, and the second is removing the routes of cross-coupling transfer.
Vibration isolation at low frequencies can be realized with an active control system and/or a spring-antispring suspension.
However, achievable displacement is ultimately limited by the noise of seismometers about $10^{-9}$ m/$\sqrt{\rm Hz}$ for the active control system \cite{AVI} since we have to know how much the ground vibration is.
For the spring-antispring system, suspension thermal noise of $10^{-15}$ m/$\sqrt{\rm Hz}$ @0.1 Hz \cite{sp-asp} will be the limitation.
In any case, removing the coupling routes is required to at least the $10^{-5}$ rad/m level.

However, there has only been limited research on cross-coupling transfer.
There is a model calculation result for a specific torsion pendulum known as TorPeDO (Torsion Pendulum Dual Oscillator), which showed that the value of $10^{-2}$ rad/m is feasible \cite{TorPeDO}.
To better understand how to reduce the couplings, we have to calculate the coupling transfer functions for more general cases.
But first we need to identify the routes of cross-coupling.

In this paper we show how seismic cross-coupling noise is introduced and how they can be removed in a standard torsion pendulum.
Routes of cross-coupling transfer are investigated and calculated in Sec.\ref{sec:scc}.
We then experimentally measured cross-coupling transfer functions in a two-stage torsion pendulum and reduced them, which is reported in Sec.\ref{sec:exp}.

\section{\label{sec:scc}Theory of Seismic Cross-coupling}

\begin{figure}
\includegraphics[width=7cm]{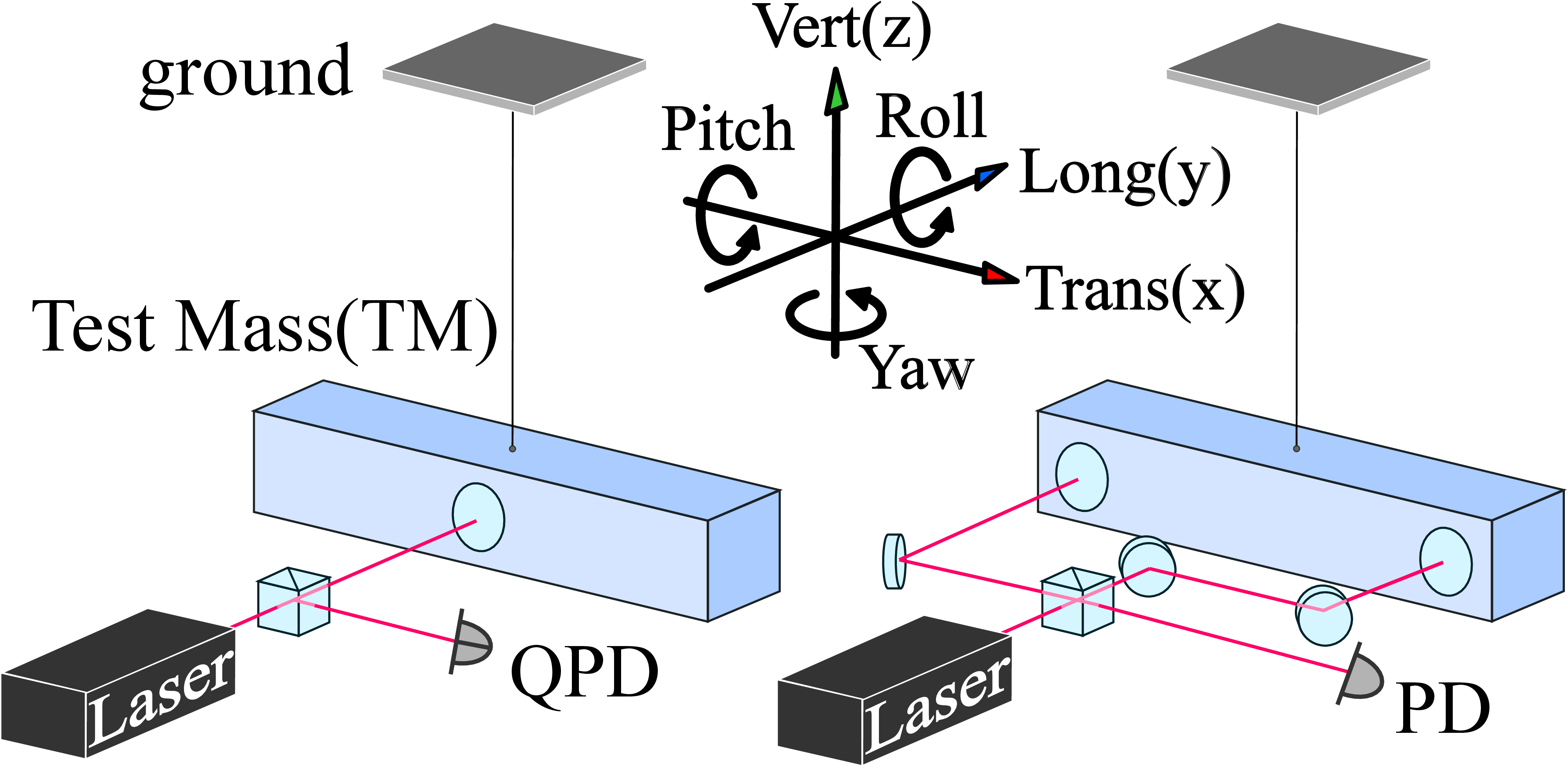}
\colorcaption{\label{fig:model}Model and coordinates. An optical lever (left) and Michelson interferometer (right).}
\end{figure}

\renewcommand{\arraystretch}{1.3}
\begin{table}
	\caption{Resonant modes and frequencies of a pendulum. $l$ is the length of the suspension wire, $h$ is the vertical offset between the suspension point and the center of mass, $g$ is gravitational acceleration, $k$, $\kappa_{\rm Y}$ are the elastic spring constant of the wire for stretching and torsion respectively, $m$ is mass of the bar and $I_{\rm P(R,Y)}$ is the moment of inertia around each axis. Calculated values for the test mass of current prototype TOBA are also shown in the fourth column.}
	\begin{center}
	\begin{tabular}{cccc}\hline\hline
		mode		&	symbol	&	frequency							&	values of TOBA [Hz]	\\ \hline
		Long, Trans	&	L, T		&	$\sqrt{g/(l+h)}/2\pi$						& 	1.0	 	\\ 
		Vert			&	V		&	$\sqrt{k/m}/2\pi$						&	30		\\
		Pitch		& 	P		&	$\sqrt{mgh/I_{\rm P}}/2\pi$			&	0.3		\\
		Roll			&	R		&	$\sqrt{mgh/I_{\rm R}}/2\pi$			& 	1.4		\\	
		Yaw			&	Y		& 	$\sqrt{\kappa_{\rm Y}/I_{\rm Y}}/2\pi$	& 	0.011	\\
		\hline\hline
	\end{tabular}
	\label{table:modes}
	\end{center}
\end{table}
\renewcommand{\arraystretch}{1.0}

In the following sections, we define the coordinates as shown in Fig. \ref{fig:model}.
Each symbol and each associated frequency are shown in Table \ref{table:modes}.
The three translational degrees of freedom, longitudinal, transverse and vertical, will be shortened as Long, Trans and Vert, respectively.
The rotation around the Vert axis is called Yaw, which is the main motion we want to measure.
Translational ground motion will not be transferred to the Yaw rotational signal in an ideally symmetric torsion pendulum.
In a realistic system however, asymmetry of the system can lead to cross-coupling.

First, we need to define the angular sensor for a torsion pendulum in order to discuss about what kind of coupling signal exists because some coupling routes depend on the sensing system.
For angular sensing, optical levers or interferometric sensors are typically used (Fig. \ref{fig:model}). 
An optical lever has been used for the test of gravitational inverse-square law \cite{Washington2}, and interferometric angular sensors are used in TOBA, seismic tiltmeters and other systems \cite{Phase1TOBA, NNcancel_tiltmeter, AngularSensor_FEPA}.
Both type of sensors measure the direction of the normal vector of the mass surface with a reflected laser beam, though the configurations are different.
Horizontal and vertical components of the normal vector are generated by rotation of the mass, and the horizontal component contains the Yaw signal which we want to measure.

In this case, the seismic cross-coupling can be classified into the following three types; (i) dynamical transfer to actual Yaw rotation, (ii) horizontal component of the normal vector generated by other rotations and translations, and (iii) vertical component of the normal vector detected via the asymmetry of the sensor.
Type (i) exists in all torsion pendulums, while (ii) and (iii) depend on the sensing system, but are similar in optical levers and interferometric sensors.
The details of these are discussed in the following subsections.

\subsection{\label{sec:scc:motion}Dynamical transfer to Yaw rotation}
\begin{figure}[tbhp]
\includegraphics[width=5.5cm]{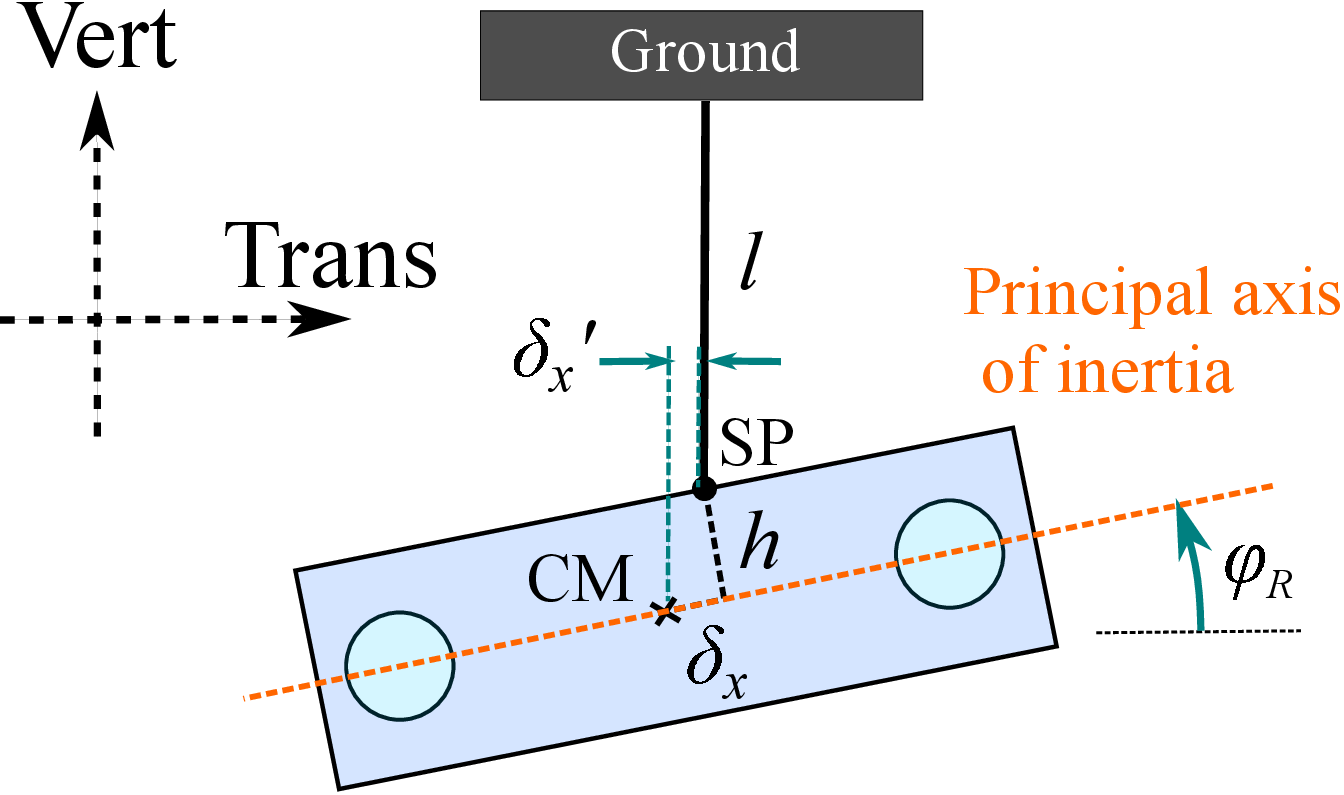}
\colorcaption{\label{fig:cc1}Mechanical asymmetry of a torsion pendulum. The SP-CM offset $\delta_x$ and the Roll of the principal axis $\varphi_{\rm R}$ are shown here.}
\end{figure}

Dynamical cross-coupling transfer can be introduced by mechanical asymmetry such as an offset between the suspension point (SP) and the center of mass (CM) as well as asymmetry of mass distribution.
These transfers can be calculated from equations of motion.
First we calculate the transfer function from ground motion in the Long direction. 
In this case, to the first order approximation, it is enough to consider the motion in Long, Pitch and Yaw direction.
Here we define the SP-CM offset along the principal axis of inertia as $\delta_x$, the SP distance from the principal axis as $h$, and the Roll tilt of the axis as $\varphi_{\rm R}$ (Fig. \ref{fig:cc1}).
We limit the discussion to the case of small asymmetry, $\delta_x/h \ll 1$ and $\varphi_{\rm R} \ll 1$, which is satisfied easily.

The Lagrangian of the system is
\begin{equation}
\begin{split}
\mathcal{L} = 
\frac{1}{2}m\dot{y}_{\rm L}{}^2 
+ \frac{1}{2}I_{\rm P}\dot{\theta}_{\rm P}{}^2 
+ \frac{1}{2}I_{\rm Y}\dot{\theta}_{\rm Y}&{}^2 
+ \varphi_{\rm R} ( I_{\rm P}-I_{\rm Y} )\dot{\theta}_{\rm P}\dot{\theta}_{\rm Y}\\
&- m g z_{\rm CM}
- \frac{1}{2} \kappa_{\rm Y} \theta_{\rm Y}{}^2,
\end{split}
\end{equation}
where $y_{\rm L}$, $\theta_{\rm P}$ and $\theta_{\rm Y}$ are the positions of the CM in Long direction, rotational angle around the CM in Pitch and Yaw direction, respectively.
$I_{\rm P},I_{\rm R}$ and $I_{\rm Y}$ are moments of inertia of the test mass (TM) around the principal axes.
$m$ is the mass of the TM, $g$ is gravitational acceleration, and $\kappa_{\rm Y}$ is the torsional spring constant of the wire.
The fourth term is the cross term of rotational energy introduced by the Roll of the principal axis.

Vert position of the CM $z_{\rm CM}$ is calculated geometrically as
\begin{equation}
z_{\rm CM} =\frac{1}{2l} \left( y_{\rm L} - y_g - h^\prime \theta_{\rm P} + \delta_x^\prime \theta_{\rm Y} \right)^2 + \frac{1}{2} h^\prime \theta_{\rm P}{}^2 - \delta_x^\prime \theta_{\rm P}\theta_{\rm Y}. \label{eq:zcm}
\end{equation}
Here $\delta_x^\prime\equiv(\delta_x - \varphi_{\rm R} h)$ and $h^\prime\equiv(h+\varphi_{\rm R}\delta_x )\simeq h$ are the relative positions between the SP and the CM in Trans and Vert directions, respectively. $y_g$ is the motion of the ground in Long direction and $l$ is the length of the suspension wire. 

By solving the Fourier transformed Euler-Lagrange equations for $\theta_{\rm Y}$, we get
\begin{equation}
\tilde{H}_{{\rm L}\rightarrow{\rm Y}}(\omega) \equiv \frac{\tilde{\theta}_{\rm Y} }{\tilde{y}_g}
\simeq
\frac{- \left( \varphi_{\rm R} - \frac{I_{\rm P}}{I_{\rm Y}} \frac{\delta_x}{h}  \right) \frac{\omega^2}{\omega_{\rm Y}{}^2}\frac{\omega^2}{g}}{\left(1-\frac{\omega^2}{\omega_{\rm Y}{}^2}\right)\left(1-\frac{\omega^2}{\omega_{\rm L}{}^2}\right)\left(1-\frac{\omega^2}{\omega_{\rm P}{}^2}\right)}
\label{eq:HLY}
\end{equation}
to first order in asymmetry, $\varphi_{\rm R}$ and $\delta_x/h$.
This is the cross-coupling transfer function from Long motion of the ground to Yaw rotation of the TM.
Here, $\omega_{\rm L} \equiv \sqrt{g/(l+h)}$, $\omega_{\rm P} \equiv \sqrt{m g h/I_{\rm P}}$ and $\omega_{\rm Y} \equiv \sqrt{\kappa_{\rm Y}/I_{\rm Y}}$ are the resonant angular frequencies of Long, Pitch and Yaw respectively.

The coupling transfer function (\ref{eq:HLY}) is proportional to a coefficient $\varphi_{\rm R} - \frac{I_{\rm P}}{I_{\rm Y}} \frac{\delta_x}{h} \equiv \varphi_{\rm R}^\prime$. 
This can roughly be interpreted as the Roll tilt of Pitch rotational axis of the TM because the Yaw component of the motion is proportional to the Roll tilt.
The rotation of the mass tends to be close to Pitch around its principal axis which has a Roll tilt of $\varphi_{\rm R}$, but torque in the Yaw axis from the wire slightly changes the axis by the second term of $\varphi_{\rm R}^\prime$.

The transfer function from Long motion of the ground to Pitch rotation of the TM, $\tilde{H}_{{\rm L}\rightarrow{\rm P}}$, can also be derived from the Euler-Lagrange equations, it being

\begin{equation}
\tilde{H}_{{\rm L}\rightarrow{\rm P}}
\simeq
\frac{ \frac{\omega^2}{g} }{\left(1-\frac{\omega^2}{\omega_{\rm L}{}^2}\right)\left(1-\frac{\omega^2}{\omega_{\rm P}{}^2}\right)}.
\label{eq:HLR}
\end{equation}
Therefore, the coupling transfer function (\ref{eq:HLY}) can be written as
\begin{equation}
\tilde{H}_{{\rm L}\rightarrow{\rm Y}}
=
\varphi_{\rm R}^\prime
\frac{ - \omega^2 }{ \omega_{\rm Y}{}^2 - \omega^2 }
\tilde{H}_{{\rm L}\rightarrow{\rm P}}.
\label{eq:HLY2}
\end{equation}
Along the observation band of TOBA, which is above the Yaw resonant frequency, Eq. (\ref{eq:HLY2}) becomes $\tilde{H}_{{\rm L}\rightarrow{\rm Y}} \simeq \varphi_{\rm R}^\prime \tilde{H}_{{\rm L}\rightarrow{\rm P}}$.
The coupling signal is thus approximately proportional to Pitch rotation.

The coupling transfer from Trans motion of the ground can be calculated in the same manner.
Here, it is enough to consider the motion of the TM in Trans, Roll and Yaw.
If we define the SP-CM offset $\delta_y$ and the Pitch tilt of the principal axis of inertia $\varphi_{\rm P}$, a coupling transfer function from Trans of the ground to Yaw of the TM, $\tilde{H}_{{\rm T}\rightarrow{\rm Y}}$, is 
\begin{equation}
\begin{split}
\tilde{H}_{{\rm T}\rightarrow{\rm Y}}(\omega)
&\simeq
\frac{- \left( \varphi_{\rm P} - \frac{I_{\rm R}}{I_{\rm Y}} \frac{\delta_y}{h}  \right) \frac{\omega^2}{\omega_{\rm Y}{}^2}\frac{\omega^2}{g}}{\left(1-\frac{\omega^2}{\omega_{\rm Y}{}^2}\right)\left(1-\frac{\omega^2}{\omega_{\rm T}{}^2}\right)\left(1-\frac{\omega^2}{\omega_{\rm R}{}^2}\right)}\\
&=
\varphi_{\rm P}^\prime
\frac{ - \omega^2 }{ \omega_{\rm Y}{}^2 - \omega^2 }
\tilde{H}_{{\rm T}\rightarrow{\rm R}}
\label{eq:HTY}.
\end{split}
\end{equation}
$\omega_{\rm T} \equiv \sqrt{g/(l+h)}$ and $\omega_{\rm R} \equiv \sqrt{m g h/I_{\rm R}}$ are resonant angular frequencies of Trans and Roll, respectively.
The coefficient $\varphi_{\rm P}^\prime$ is the Pitch tilt of Roll rotational axis and $\tilde{H}_{{\rm T}\rightarrow{\rm R}}$ is the transfer function from Trans of the ground to Roll of the TM.

\subsection{\label{sec:scc:hori}Horizontal component of the normal vector generated by other DoFs}

\begin{figure}
\includegraphics[width=4cm]{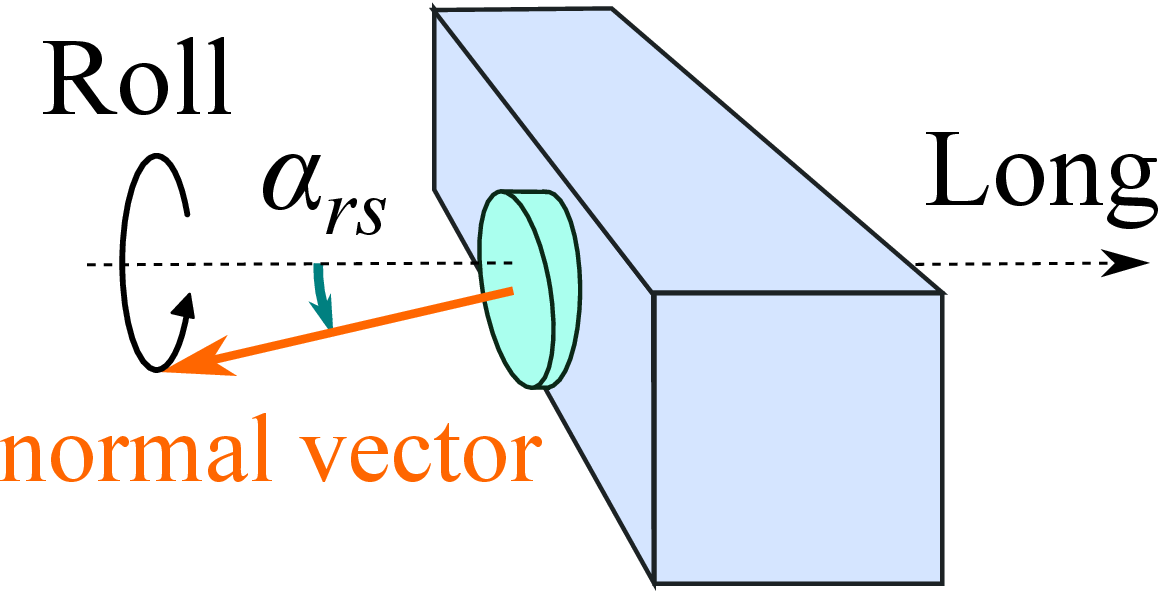}
\colorcaption{\label{fig:ccrs}Tilt of the reflecting surface. The normal vector of the surface has a Pitch tilt of $\alpha_{\rm rs}$.}
\end{figure}
First, when the reflecting surface on the TM has a Pitch tilt $\alpha_{\rm rs}$ (Fig. \ref{fig:ccrs}), Roll rotation $\theta_{\rm R}$ generates a Trans component of unit normal vector of $\sin\alpha_{\rm rs}\sin\theta_{\rm R} \simeq \alpha_{\rm rs}\theta_{\rm R}$.
This is read by the sensor as Yaw rotation, thus the coupling transfer function $\tilde{H}_{\rm c,rs}$ is
\begin{equation}
\tilde{H}_{\rm c,rs} = \alpha_{\rm rs}\tilde{H}_{{\rm T}\rightarrow{\rm R}}.
\label{eq:rs}
\end{equation}
The origin of this coupling is Trans motion of the ground.

\begin{figure}
\includegraphics[width=7cm]{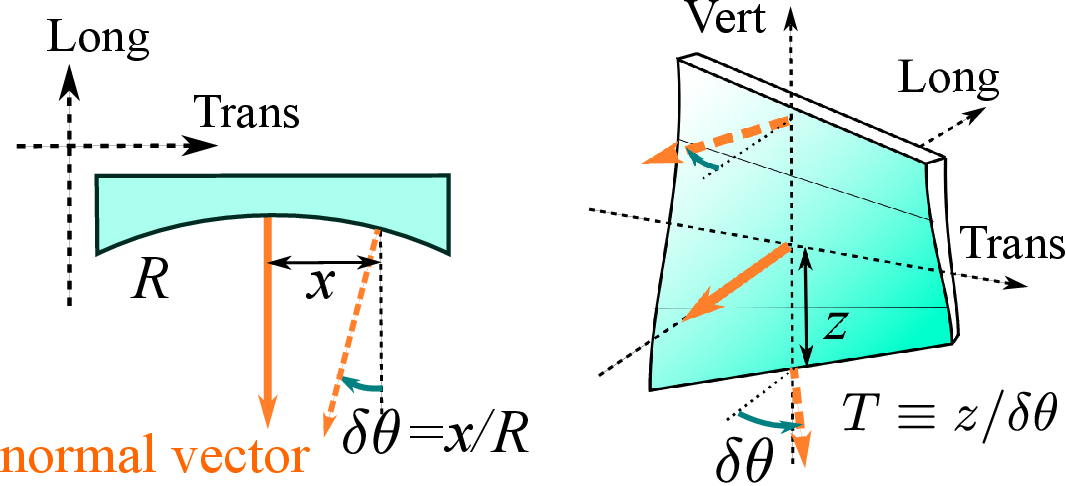}
\colorcaption{\label{fig:ccroctw}Curvature and twist of the reflecting surface. The direction of the normal vector depends on the position.}
\end{figure}
Additionally, if the reflecting surface is not flat causing the direction of the normal vector to depend on position, Trans and Vert translation of the mass will couple to the signal.
For example, if the surface has a radius of curvature $R$, Trans translation $x$ of the mass changes the direction of the normal vector by $x/R$ rad.
This would be the coupling signal.
A second example would be if the surface is a twisted shape as in Fig. \ref{fig:ccroctw}, Vert translation $z$ will couple to the signal.
In analogy to $R$, here we define $T$ as the ratio of vertical offset to angle change of the normal vector.
Coupling transfer functions for these are
\begin{equation}
\tilde{H}_{\rm c,curve} = \frac{1}{R} \tilde{H}_{{\rm T}\rightarrow{\rm T}}	\label{eq:roc}	\\
\end{equation}
and
\begin{equation}
\tilde{H}_{\rm c,twist} = \frac{1}{T} \tilde{H}_{{\rm V}\rightarrow{\rm V}}.	\label{eq:twist}
\end{equation}
$\tilde{H}_{{\rm T}\rightarrow{\rm T}}$ and $\tilde{H}_{{\rm V}\rightarrow{\rm V}}$ are the transfer functions from Long motion of the ground to Long translation of the TM, and Vert to Vert, respectively.  
Typically these are smaller than other coupling transfer functions since plane mirrors have good flatness.
For a Michelson interferometer which uses two beams separated by $L$, the equivalent $R$ and $T$ are calculated as $R=L/\phi_{\rm Y}$ and $T=L/\phi_{\rm P}$ where $\phi_{\rm Y}$ and $\phi_{\rm P}$ are the relative tilt of the surface between two beamspots in Yaw and Pitch respectively.

Other DoFs generate no horizontal component of the normal vector or are much smaller than these by orders of magnitude.

\subsection{\label{sec:scc:vert}Coupling from vertical component of the normal vector}
\begin{figure}
\includegraphics[width=7cm]{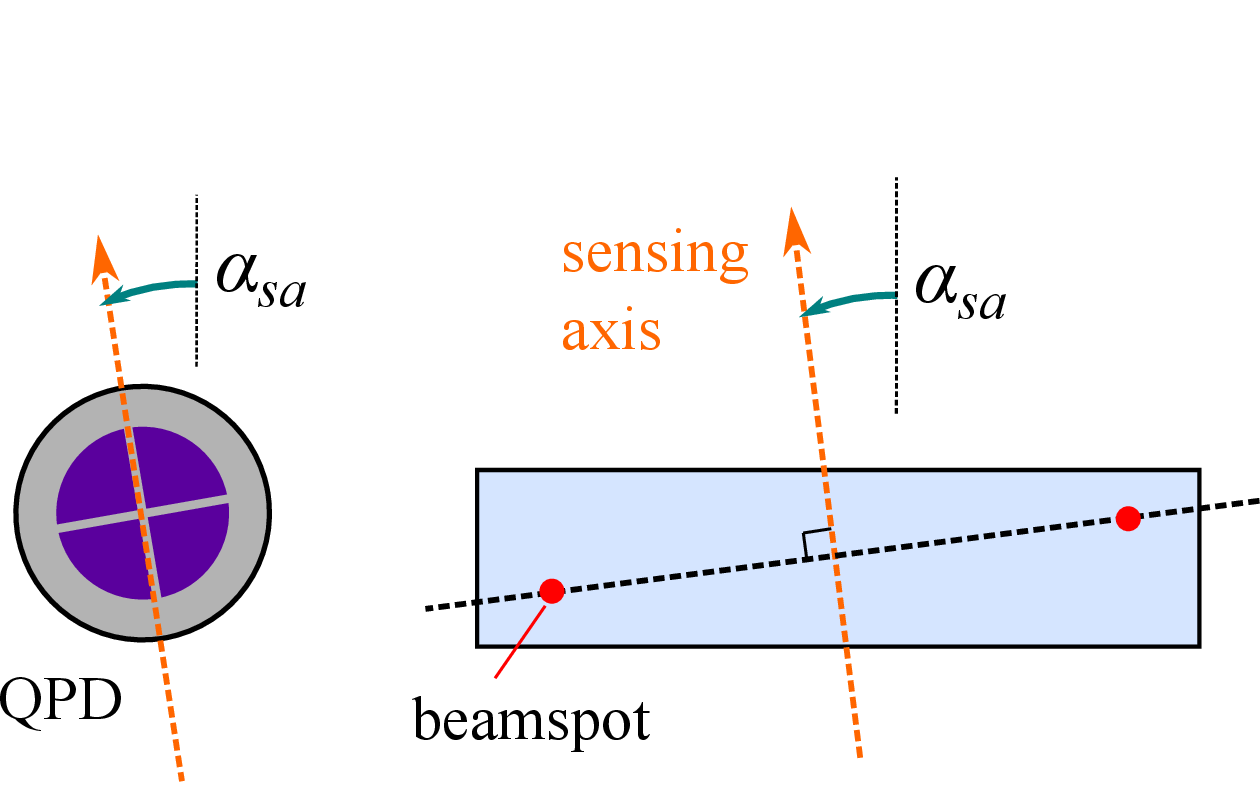}
\colorcaption{\label{fig:ccsa}Sensing axes defined by the direction of the QPD (left) or two beamspots (right).}
\end{figure}
Next we calculate the coupling from vertical component of the normal vector.
Most of angular sensors measure rotation of a target mass around a certain axis, which we call the ``sensing axis'' here.
For example, a sensing axis is determined by the direction of the QPD for an optical lever and by two beamspots for a Michelson interferometer (Fig. \ref{fig:ccsa}).
When the sensing axis tilts in Roll by $\alpha_{\rm sa}$, its readout signal is $\cos\alpha_{\rm sa}\times({\rm Yaw}) - \sin\alpha_{\rm sa}\times({\rm Pitch}) \simeq ({\rm Yaw}) - \alpha_{\rm sa} ({\rm Pitch})$, with the second term being the coupling signal.
Thus the coupling transfer function $\tilde{H}_{\rm c,sa}$ is
\begin{equation}
\tilde{H}_{\rm c,sa} = - \alpha_{\rm sa} \tilde{H}_{{\rm L}\rightarrow{\rm P}}.
\label{eq:sa}
\end{equation}
This cross-coupling is caused by Long motion of the ground.

\subsection{\label{sec:scc:total}Total transfer function}
Coupling transfer functions for each coupling route have been calculated.
Here we describe the sum of them, which actually appears in experiments.

First, the total coupling transfer function from Long motion of the ground, $\tilde{H}_{c,{\rm Long}}$, is the sum of (\ref{eq:HLY2}) and (\ref{eq:sa}), given by
\begin{eqnarray}
\tilde{H}_{c,{\rm Long}} &=& 
\left( \varphi_{\rm R}^\prime \frac{ - \omega^2 }{ \omega_{\rm Y}{}^2 - \omega^2 } - \alpha_{\rm sa} \right)
\tilde{H}_{{\rm L}\rightarrow{\rm P}} \nonumber\\
&\simeq& 
\left( \varphi_{\rm R}^\prime - \alpha_{\rm sa} \right) \tilde{H}_{{\rm L}\rightarrow{\rm P}}.	\label{eq:HcL}
\end{eqnarray}
The approximation is valid when above the Yaw resonant frequency.

Total coupling transfer functions from Trans motion, $\tilde{H}_{c,{\rm Trans}}$, is the sum of (\ref{eq:HTY}), (\ref{eq:rs}) and (\ref{eq:roc}), and is
\begin{eqnarray}
\tilde{H}_{c,{\rm Trans}} &=& 
\left( \varphi_{\rm P}^\prime \frac{ - \omega^2 }{ \omega_{\rm Y}{}^2 - \omega^2 } + \alpha_{\rm rs} \right)
\tilde{H}_{{\rm T}\rightarrow{\rm R}} 
+ \frac{1}{R} \tilde{H}_{{\rm T}\rightarrow{\rm T}} \nonumber\\
&\simeq& 
\left( \varphi_{\rm P}^\prime + \alpha_{\rm rs} \right) \tilde{H}_{{\rm T}\rightarrow{\rm R}}
+\frac{1}{R} \tilde{H}_{{\rm T}\rightarrow{\rm T}}.	\label{eq:HcT}
\end{eqnarray}

There is only one route of cross-coupling from Vert, so the total coupling transfer functions from Vert, $\tilde{H}_{c,{\rm Vert}}$, is (\ref{eq:twist}) itself.

These cross-coupling transfer functions can apparently be reduced by adjusting the coefficients $(\varphi_{\rm R}^\prime - \alpha_{\rm sa})$,  $(\varphi_{\rm P}^\prime + \alpha_{\rm rs})$, R and T, with the first two representing the Roll and Pitch tilt of the system and the latter two the flatness of the reflecting surface.
We can calculate these coefficients by measurements of the cross-coupling transfer functions.
Suppressing the motion of the mass, $\tilde{H}_{{\rm L}\rightarrow{\rm P}}$ and so on, is also effective in reducing coupling.
Off-line subtraction by monitoring all DoFs with auxiliary sensors is also possible.

\section{\label{sec:exp}Measurement and reduction}
In this section, we report experimental results about measurement and reduction of cross-coupling transfer functions in order to demonstrate our calculations above. 
We measured the coupling from Long and Trans motion of the ground, as they are the most important contributions for TOBA since the associated resonant frequencies are close to its observation band (Table \ref{table:modes}).

\subsection{\label{sec:exp:setup}Setup}
\begin{figure}
\includegraphics[width=7cm]{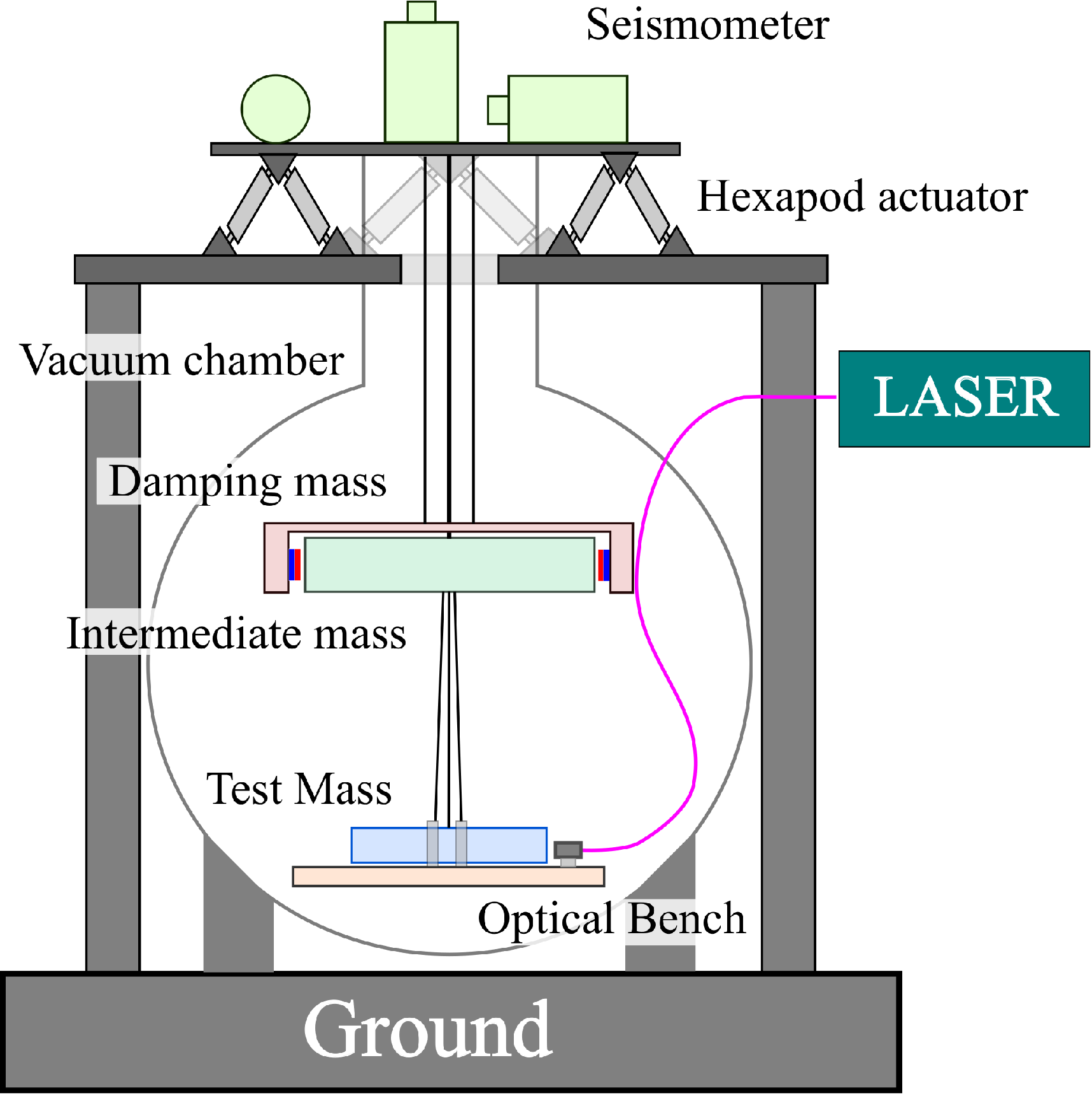}
\colorcaption{\label{fig:setup}Experimental setup.}
\end{figure}
An overview of the setup is shown in Fig. \ref{fig:setup}.
We used a double-stage torsion pendulum.
The test mass (TM) is a bar shaped fused silica of 20 cm $\times$ 3 cm $\times$ 3 cm.
One surface of the TM is polished and coated to reduce the relative Pitch and Yaw of the surface between the two beamspots.
The Michelson interferometer used to read Yaw rotation is constructed on the optical bench (OB).
The TM and the OB are suspended from the intermediate mass, and the intermediate mass is suspended from the top plate.
The intermediate mass is damped by magnets on the damping mass in order to suppress resonance of the TM and the OB.
Laser light is sent from the source to the OB through an optical fiber, whose midpoint is attached to the damping mass in order to reduce vibration transfer via the fiber.
At the top stage, hexapod actuators are set to shake the suspension point.
Six seismometers are placed on the stage to measure the vibration of the suspension point.
Coupling transfer functions, from seismometer signal to interferometer signal, are measured by shaking the hexapod actuators.

\subsection{\label{sec:exp:meas}Transfer function measurement}
Cross-coupling transfer can occur for both TM and OB in our setup, so the theoretical transfer functions to be measured are modified from Eq. (\ref{eq:HcL}) and (\ref{eq:HcT}).
In the following sections, we use $\tau$ and $\beta$ to express the tilt of the TM and the OB respectively instead of $\varphi$.
The transfer functions $\tilde{H}$ are also replaced with $\tilde{T}$ and $\tilde{B}$ for the TM and the OB, respectively.
First, the coupling transfer function from Long motion is modified to 
\begin{eqnarray}
\tilde{H}_{c,{\rm Long}} \simeq
\left( \tau_{\rm R}^\prime - \alpha_{\rm sa} \right) \tilde{T}_{{\rm L}\rightarrow{\rm P}} - \left( \beta_{\rm R}^\prime - \alpha_{\rm sa} \right) \tilde{B}_{{\rm L}\rightarrow{\rm P}},
\label{eq:HcL:m}
\end{eqnarray}
due to coupling transfer to the OB.
In the same manner, the coupling transfer function from Trans motion is also modified to
\begin{eqnarray}
\tilde{H}_{c,{\rm Trans}} \simeq
\left( \tau_{\rm P}^\prime + \alpha_{\rm rs} \right) \tilde{T}_{{\rm T}\rightarrow{\rm R}} -\left( \beta_{\rm P}^\prime + \alpha_{\rm rs} \right) \tilde{B}_{{\rm T}\rightarrow{\rm R}} .
\label{eq:HcT:m}
\end{eqnarray}
The third term of Eq. (\ref{eq:HcT}), the contribution of relative Yaw of the mirror surface, is neglected here since the polished surface of the TM is expected to have sufficiently small relative angle, with an order of $10^{-6}$ rad, in our experiment.
This value is roughly estimated from the surface quality $\lambda/10\sim10^{-7}$ m divided by the length of the bar $\sim 20$ cm.

We use Eq. (\ref{eq:HcL:m}) and (\ref{eq:HcT:m}) to analyze the measured coupling transfer functions.
The transfer functions $\tilde{T}_{{\rm L}\rightarrow{\rm P}}$, $\tilde{T}_{{\rm T}\rightarrow{\rm R}}$, $\tilde{B}_{{\rm L}\rightarrow{\rm P}}$ and $\tilde{B}_{{\rm T}\rightarrow{\rm R}}$ are calculated for the TM and the OB by solving the dynamical equations of motions of the whole suspension system.
Using them, the coefficients of each term can be measured by fitting the data to the equations.
The measured coefficients are used for reduction of cross-coupling because they are associated with the Roll and the Pitch of the TM and the OB.
This is our basic cross-coupling reduction strategy.
We show the experimental demonstration of this in the following subsections.

\subsection{\label{sec:exp:result}Result and analysis}

\begin{figure}
	\begin{tabular}{c}
	\begin{minipage}{1\hsize}
	\begin{center}
	\includegraphics[width=8.5cm]{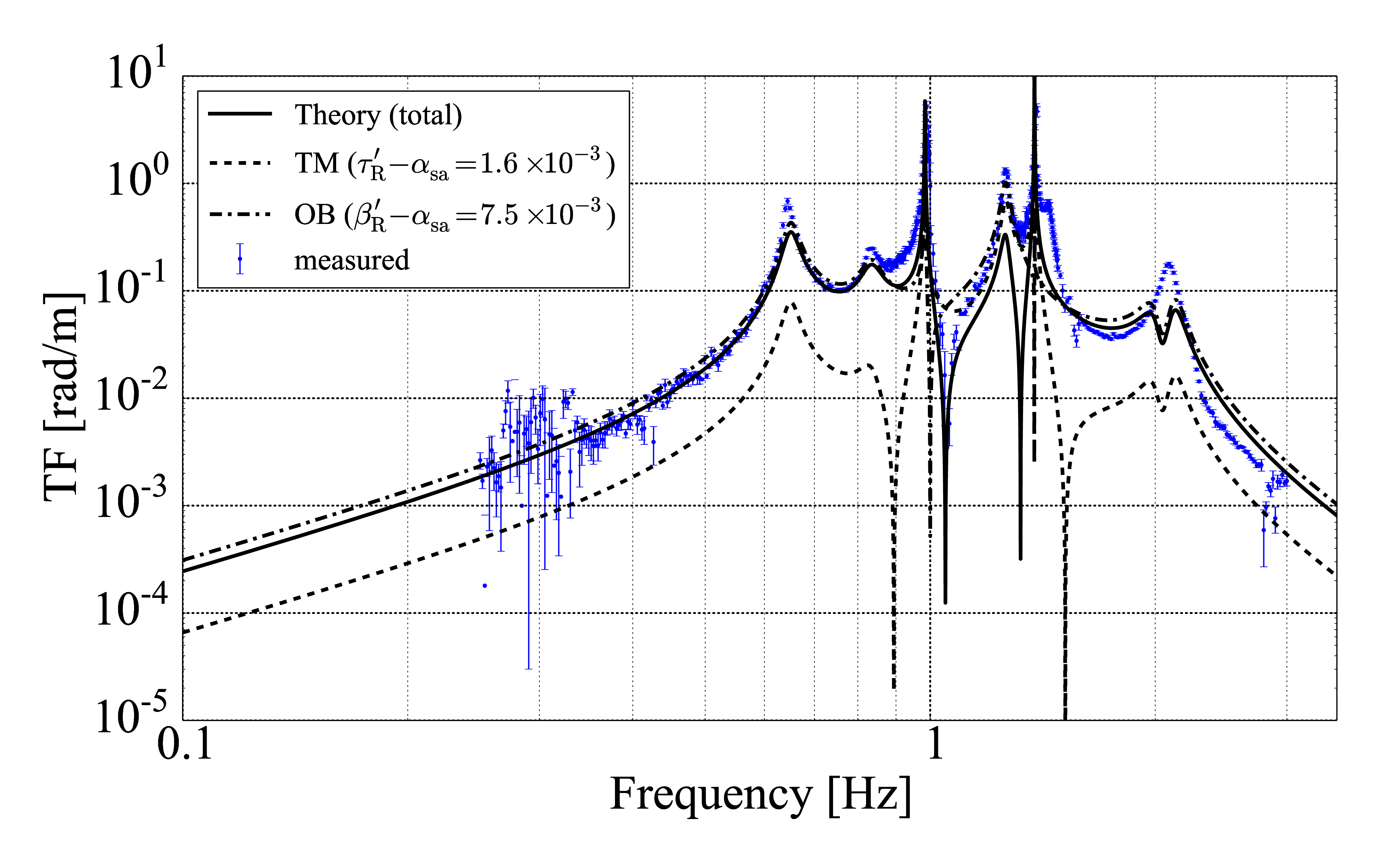}
	\end{center}
	\end{minipage}\\
	\begin{minipage}{1\hsize}
	\begin{center}
	\includegraphics[width=8.5cm]{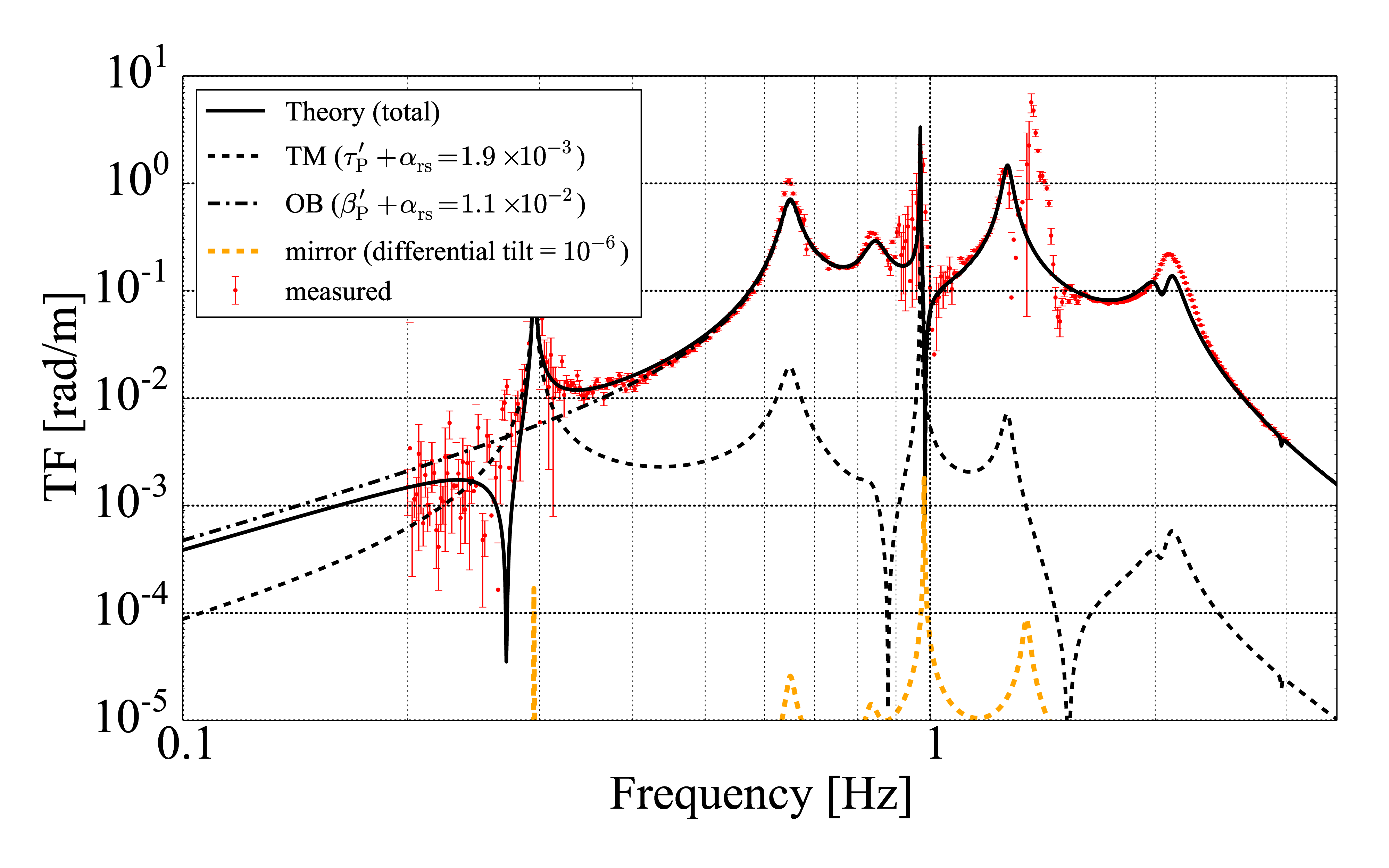}
	\end{center}
	\end{minipage}
	\end{tabular}
	\colorcaption{\label{fig:cctf0}(top) Measured cross-coupling transfer function in the Long direction. Blue dots show measured data. Contributions of the TM and the OB are also plotted with a black dashed line and a dot-dash line, respectively, and the black thick line is the sum of both. (bottom) Measured cross-coupling transfer function in the Trans direction. Red dots are measured data, and black lines are the same as Long but for the Trans direction. The orange line shows the estimated contribution from the relative Yawz of the mirror surface, when $\phi_{\rm Y}=10^{-6}$ rad.}
\end{figure}

Measured coupling transfer functions are shown in Fig. \ref{fig:cctf0}. 
The measurements were performed above 0.25 Hz and 0.2Hz for Long and Trans respectively because of larger instrumental noise at lower frequencies. 
They show good agreement with theoretical transfer functions described by Eq. (\ref{eq:HcL:m}) and (\ref{eq:HcT:m}), which are shown with black lines in the graph.
In our measurements, the transfer functions of two directions can be mixed up to 10\% due to the uncertainty of Yaw angle of the suspension system which drifts during evacuation.
The coordinates based on the pendulum are different from the ones based on the seismometers, hence when we shake the hexapod in the seismometer-based Long direction, the pendulum-based Trans vibration partially mixes.
This mixing is negligible since two transfer functions are almost the same order.

Fitted parameters were $(\tau_{\rm R}^\prime - \alpha_{\rm sa} ) = 1.6^{+4.0}_{-0.6} \times 10^{-3}$ rad and $(\beta_{\rm R}^\prime - \alpha_{\rm sa}) = (7.5\pm2.3) \times 10^{-3}$ rad for Long, and $(\tau_{\rm P}^\prime + \alpha_{\rm rs}) = (1.9\pm0.1)\times10^{-3}$ rad and $(\beta_{\rm P}^\prime + \alpha_{\rm rs}) = (1.14\pm0.05)\times10^{-2}$ rad for Trans.
Errors are mainly from systematic errors from the fitting method.
As we stated in the previous section, these coefficients indicate the Roll and the Pitch of the TM and the OB.
Both coupling transfer functions in the Long and Trans directions are dominated by the contribution of the OB.
The estimated Roll and Pitch of the OB is large, in the order of $10^{-2}$ rad, because the OB contains many optical components and thus has a large asymmetrical mass distribution, which results in a large tilt of the principal axes of inertia.
Tilt of the TM is about $10^{-3}$ rad, which corresponds to about 1 $\mu$m offset of the center of mass, which is reasonable for the initial asymmetry.

Theoretical transfer functions of the Pitch and Roll rotations, $\tilde{H}_{{\rm L}\rightarrow{\rm P}}$ and $\tilde{H}_{{\rm T}\rightarrow{\rm R}}$, are calculated with a rigid body model.
The shape of some resonant peaks do not perfectly correspond to the measured transfer functions, which may come from errors with the parameters used for the model.
At the bottom of Fig. \ref{fig:cctf0}, expected contribution of the differential mirror Yaw angle $\phi_{\rm Y}$ is shown with an orange dashed line.

\subsection{\label{sec:exp:reduce}Reduction of coupling transfer functions}
As the coupling coefficients of the TM and the OB are estimated by the previous measurement, we now know how much the Roll and Pitch should be changed to reduce cross-coupling.
They can be adjusted by using counter weights or actuators.

\begin{figure}
	\begin{tabular}{c}
	\begin{minipage}{1\hsize}
	\begin{center}
	\includegraphics[width=8.5cm]{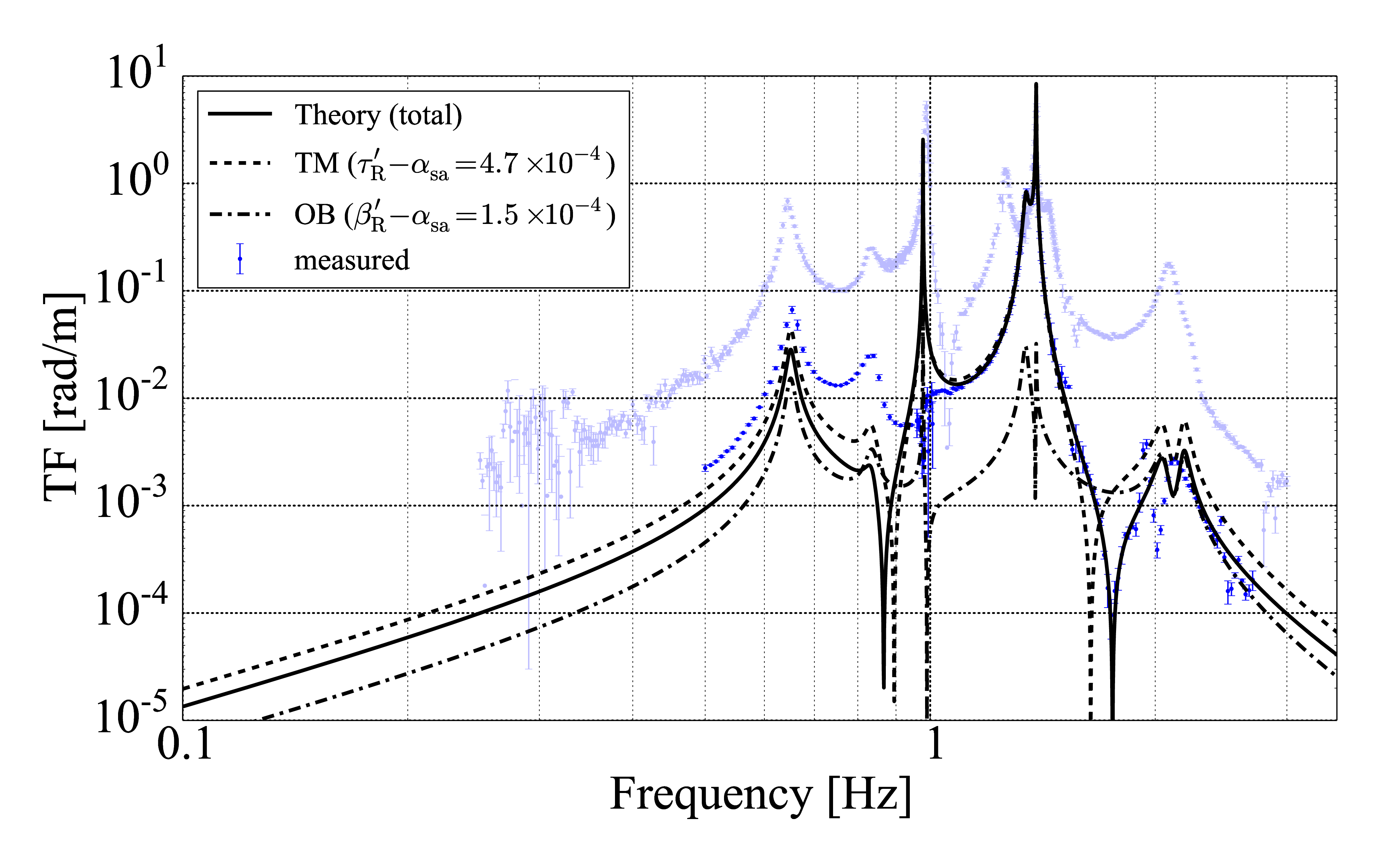}
	\end{center}
	\end{minipage}\\
	\begin{minipage}{1\hsize}
	\begin{center}
	\includegraphics[width=8.5cm]{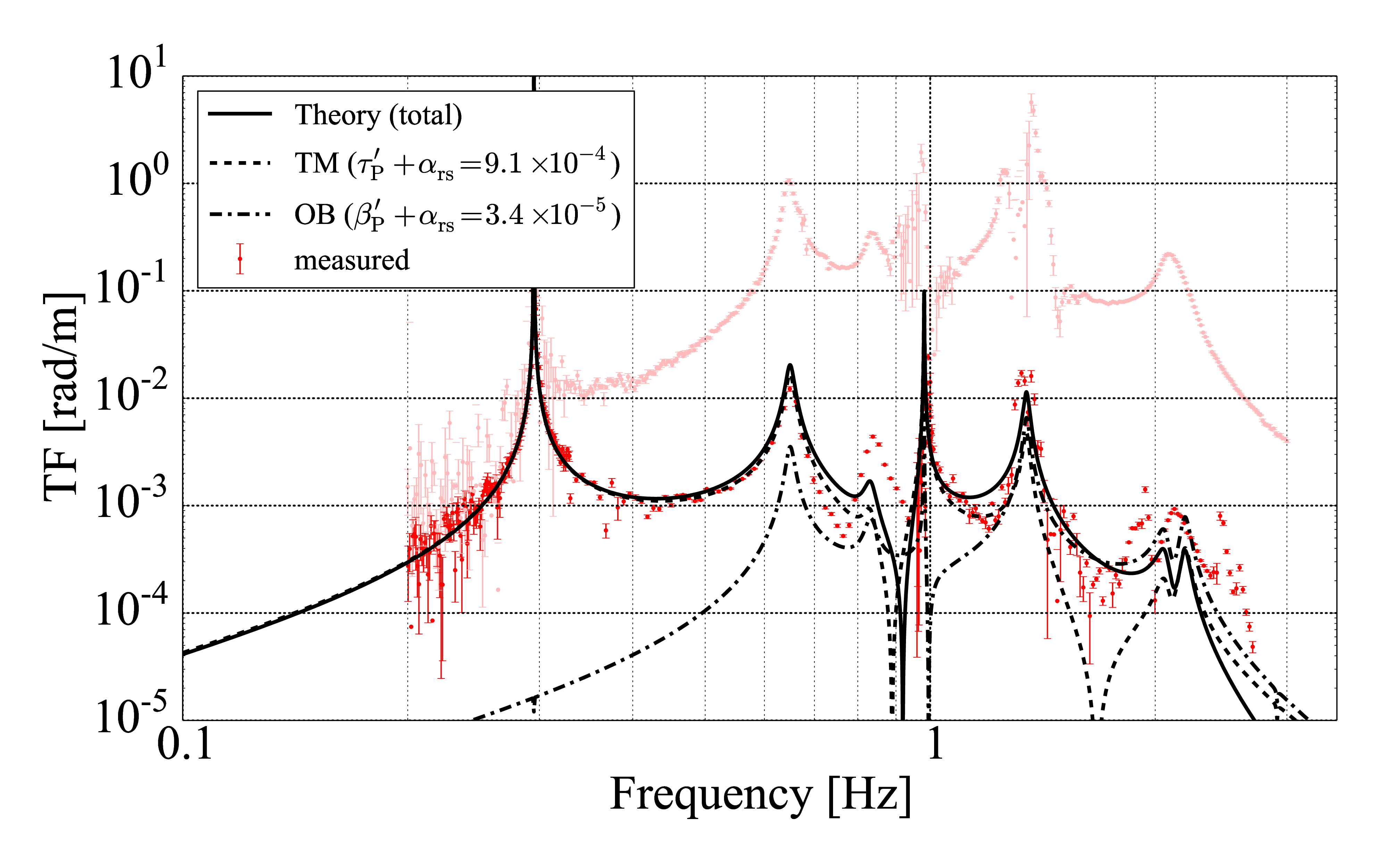}
	\end{center}
	\end{minipage}
	\end{tabular}
	\colorcaption{\label{fig:cctf1}(top) Reduced cross-coupling transfer function in Long direction (blue dots). The faint blue dots are the measured values before reduction. (bottom) Reduced cross-coupling transfer function in Trans direction (red dots). The faint red dots are the values before reduction.}
\end{figure}

\subsubsection{Reduction with counter weights}
First we put counter weights on the TM and the OB to reduce cross-coupling. 
A counter weight will change the parameters $\varphi_{\rm R}^\prime$ and $\varphi_{\rm P}^\prime$ as follows, with
\begin{eqnarray}
\Delta \varphi_{\rm R}^\prime &=& - \frac{m_{\rm cw}}{m}\frac{x}{h} + \frac{m_{\rm cw}xz}{I_{\rm Y} - I_{\rm P}} -\frac{I_{\rm P}}{I_{\rm Y}}  \frac{m_{\rm cw}}{m} \frac{x}{h}  \label{eq:dphiP}\hspace{10pt}\\
\Delta \varphi_{\rm P}^\prime &=& - \frac{m_{\rm cw}}{m}\frac{y}{h} + \frac{m_{\rm cw}yz}{I_{\rm Y} - I_{\rm R}} -\frac{I_{\rm R}}{I_{\rm Y}}  \frac{m_{\rm cw}}{m} \frac{y}{h}  \label{eq:dphiR}.\hspace{10pt}
\end{eqnarray}
$m_{\rm cw},x,y,z$ are the mass and position of the counter weight.
The first terms are geometrical, consisting of the Roll and the Pitch tilt of the mass shape due to the offset of the center of mass, and the second terms are the tilt of the principal axes relative to the shape because of the change of mass distribution, so they contribute to a change in the principal axes, the first term of $\varphi_{\rm R}^\prime = \varphi_{\rm R} - \frac{I_{\rm P}}{I_{\rm Y}} \frac{\delta_x}{h}$.
The third term comes from the offset of the center of mass, which corresponds to the second term of $\tau_{\rm R}^\prime$.
Other parameters $\alpha_{\rm sa}$ and $\alpha_{\rm rs}$ are associated with the Roll of the OB and the Pitch of the TM respectively, and given by
\begin{eqnarray}
\Delta \alpha_{\rm sa} = - \frac{m_{\rm cw}}{m_{\rm OB}}\frac{x}{h_{\rm OB}}\\
\Delta \alpha_{\rm rs} = - \frac{m_{\rm cw}}{m_{\rm TM}}\frac{x}{h_{\rm TM}}.
\end{eqnarray}

After putting some counter weights based on these equations, the cross-coupling transfer functions were successfully reduced.
The results are shown in Fig. \ref{fig:cctf1}.
Yet again, we see general agreement with theory, although there are some deviations around 0.85 Hz. 
The peak at 0.85 Hz is due to resonance of the damping mass, so there might be additional transfer via the optical fiber and lead wires connected between the OB and the damping mass that is not accounted for in theory.
Fitted parameters after the reduction are $(\tau_{\rm R}^\prime - \alpha_{\rm sa} ) = (4.73\pm0.02) \times 10^{-4}$ rad and $(\beta_{\rm R}^\prime - \alpha_{\rm sa}) = (1.49\pm0.02) \times 10^{-4}$ rad for Long, and $(\tau_{\rm P}^\prime + \alpha_{\rm rs}) = (9.09\pm0.04)\times10^{-4}$ rad and $(\beta_{\rm P}^\prime + \alpha_{\rm rs}) = (3.4\pm2.2)\times10^{-5}$ rad for Trans.
We achieved a total cross-coupling transfer of $2\times10^{-5}$ rad/m for Long and $4\times10^{-5}$ rad/m for Trans at 0.1 Hz (extrapolated from measurement).
The coupling coefficients of the OB are greatly suppressed by a factor of $\sim 1/100$.
For the TM, however, the reduction factors are only $1/3$ for Long and $1/2$ for Trans.
This is mainly because the initial asymmetry was much smaller than for the OB, and partly because the TM is smaller and thus more difficult to tune than the OB.
Especially for the Pitch tilt, the TM has almost the same moment of inertia for Yaw and Roll, hence the second term of Eq. (\ref{eq:dphiR}) changes drastically with a small change in the counter weight.

Since we can also estimate the change of tilt from the mass and position of the counter weights, we compared them with the fitted parameters to confirm that the reduction agrees with theory.
The results are shown in Table.\ref{table:reduction}. 
For the calculated errors, we assumed a position error of $\pm$0.5 mm and a mass error of $\pm$0.2 mg for the counter weights on the TM, and a position error of $\pm$2 mm and a mass error of $\pm$1 g for the OB. 
All the coupling parameter changes correspond to the expected values from calculation within the error range.
The large calculation error of $(\tau_{\rm P}^\prime + \alpha_{\rm rs})$ comes from $I_{\rm Y(TM)} \simeq I_{\rm R(TM)}$, as mentioned earlier.

\renewcommand{\arraystretch}{1.3}
\begin{table}
	\caption{Comparison of reduced amount of coupling coefficients between fitted values from measurement vs expected values from calculation.}
	\begin{center}
	\begin{tabular}{crr}\hline\hline
		Parameter	&	Measured [rad]$\hspace{6pt}$	&	Calculated [rad]$\hspace{6pt}$	\\ \hline
		$\tau_{\rm R}^\prime - \alpha_{\rm sa}$	&	$(-1.1^{+4.0}_{-0.6})\times10^{-3}$	&	$\hspace{5pt}$$(-0.71\pm0.07)\times10^{-3}$	\\ 
		$\beta_{\rm R}^\prime - \alpha_{\rm sa}$	&	$(7.4\pm2.3)\times10^{-3}$	&	$(8.9\pm0.2)\times10^{-3}$	\\ 
		$\tau_{\rm P}^\prime + \alpha_{\rm rs}$	&	$\hspace{5pt}$$(-1.0\pm0.1)\times10^{-3}$	&	$(0.1\pm1.4)\times10^{-3}$	\\ 
		$\beta_{\rm P}^\prime + \alpha_{\rm rs}$	&	$(1.1\pm0.1)\times10^{-2}$	&	$(1.08\pm0.03)\times10^{-2}$	\\	\hline\hline
	\end{tabular}
	\label{table:reduction}
	\end{center}
\end{table}
\renewcommand{\arraystretch}{1.0}

\subsubsection{Reduction with actuators}
\begin{figure}
\includegraphics[width=8.5cm]{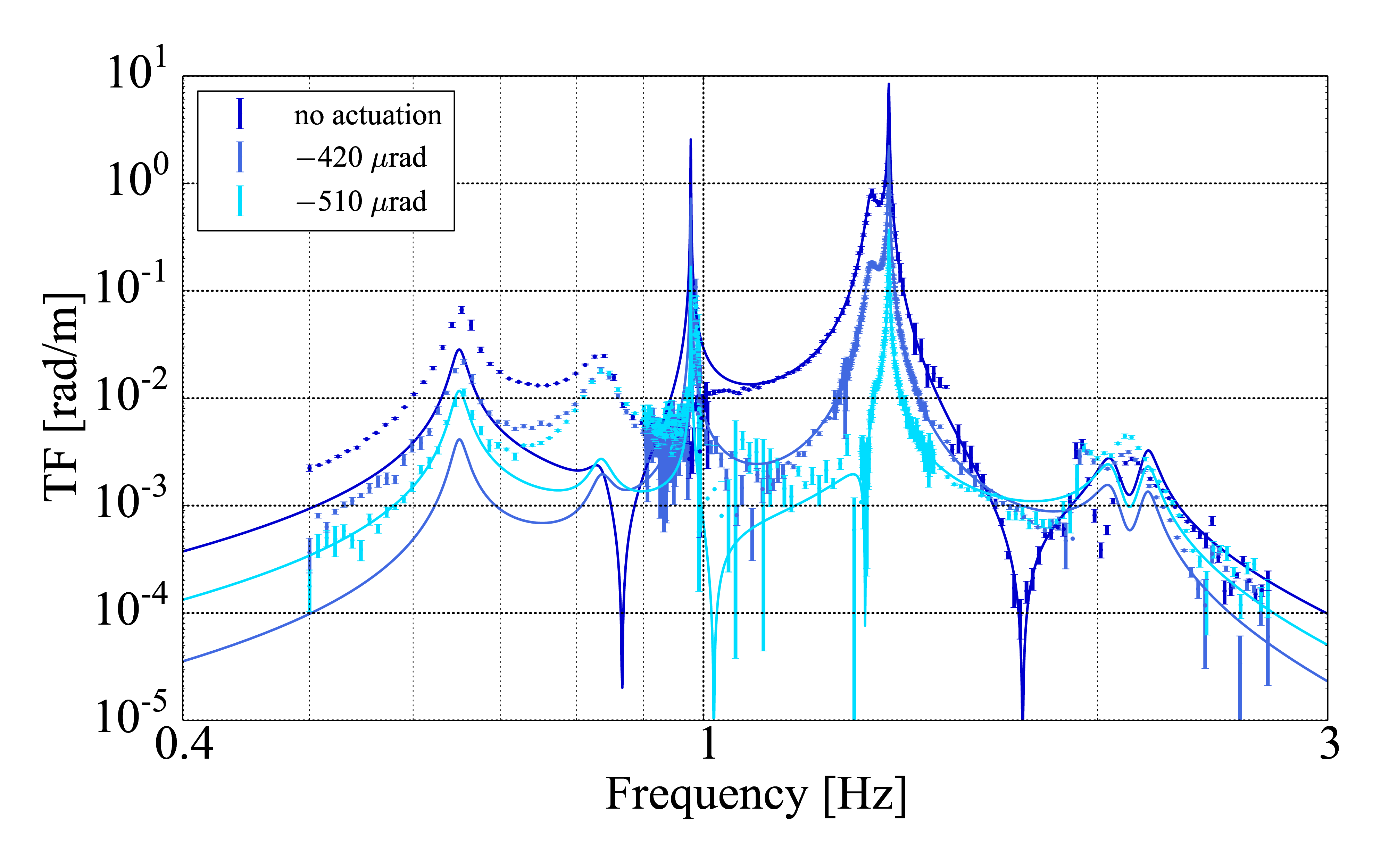}
\colorcaption{\label{fig:cctfa}Coupling transfer functions in Long as the Roll tilt of the TM is changed. Dots and solid lines are measured data and theoretical transfer function respectively, for the initial position (dark blue), $-420\,\mu$rad (blue), and $-510\,\mu$rad (light blue).}
\end{figure}
\begin{figure}
\includegraphics[width=8.5cm]{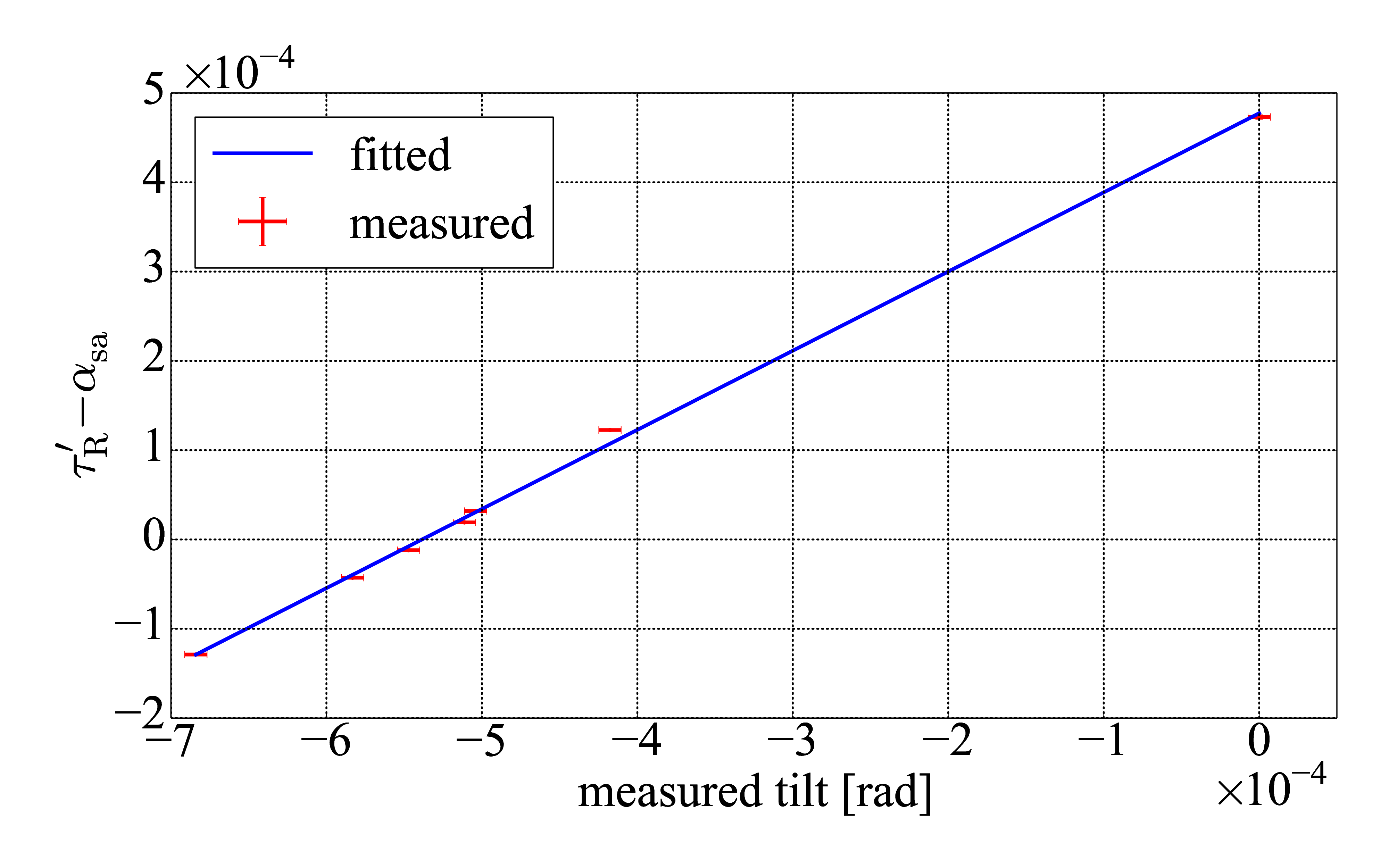}
\colorcaption{\label{fig:ccc}Fitted coupling parameter $(\tau_{\protect\rm R}^\prime - \alpha_{\rm sa} )$ vs measured Roll tilt. The proportional factor is $0.9\pm0.2$.}
\end{figure}
Next, we used actuators to reduce cross-coupling in the Long direction.
Two actuators are set at both ends of the TM and push vertically in order to change Roll tilt of the TM.
This changes the parameter $\tau_{\rm R}^\prime$ directly without affecting any other parameters.
We adjusted the Roll tilt of the TM gradually, and measured coupling transfer functions at each position.
The Roll tilt of the TM relative to the initial position was monitored with position sensors during the measurements.
Measurements were done at seven points, with three of them shown in Fig. \ref{fig:cctfa}.
The dots are the measured data and the solid lines show theoretical lines. 
Different colors indicate different Roll tilt of the TM, with the dark blue one representing the initial position (the same data as Fig. \ref{fig:cctf1}).
Cross-coupling transfer function was successfully reduced, especially at around 1.4 Hz where contribution from the TM was dominant.
Since we only changed the Roll of the TM, contribution by the OB still remained in the other frequency bands.
By extrapolating the measurement, the total coupling transfer at 0.1 Hz was estimated to be $5\times10^{-6}$ rad/m.
This is dominated by the contribution from the OB, while the contribution of the TM is suppressed to $5\times10^{-7}$ rad/m. 

We compared the fitted parameters $(\tau_{\rm R}^\prime - \alpha_{\rm sa})$ and the Roll tilt measured by position sensors (Fig. \ref{fig:ccc}).
Since $\tau_{\rm R}^\prime$ is the Roll tilt of the TM itself, it should be linear to the measured tilt with a proportional factor of 1.
The result in Fig. \ref{fig:ccc} shows a linear relation with a fitted proportional factor of $0.9\pm0.2$.
This error comes from the 20\% calibration uncertainty of the interferometer, hence a systematic error of $(\tau_{\rm R}^\prime - \alpha_{\rm sa})$.
In conclusion, reduction of coupling coefficients with actuators was also achieved in line with theory.

\section{Discussion}
We achieved cross-coupling transfer of $5\times10^{-6}$ rad/m for Long and $4\times10^{-5}$ rad/m for Trans at 0.1 Hz.
These values are close to the minimum requirement for TOBA when translation of the suspension point is suppressed to $10^{-15}$ m/$\sqrt{\rm Hz}$, the suspension thermal noise level of a spring-antispring vibration isolation system.
Additionally, if the suspension point is actively stabilized to $10^{-9}$ m/$\sqrt{\rm Hz}$, strain equivalent noise reaches $10^{-14}$ /$\sqrt{\rm Hz}$ with a $5\times10^{-6}$ rad/m cross-coupling.
This means that the achieved cross-coupling is an important milestone for geophysical purposes because $10^{-14}$ /$\sqrt{\rm Hz}$ sensitivity is close to expected Newtonian noise levels as well as estimated earthquake gravity signals.
In any case, it is technically not easy to construct such vibration isolation systems.
Therefore, cross-coupling transfer should be reduced as much as possible to relax the requirement on them.

Current cross-coupling level of the TM in Long is determined by the precision of Roll tilt adjustment, $\sim 10\,\mu$ rad. 
The other terms we considered here, cross-coupling of the TM in Trans and contributions of the OB, will be limited by the similar precision after finer tuning.
This is almost the same as the RMS amplitude of Roll/Pitch vibration of the TM and the OB which are also few tens of microradians in our case.
Hence, vibration has to be suppressed by stronger damping or feedback control for more precise adjustments.
Feedback control is the better option because strong damping for Roll/Pitch tilt can also damp Yaw rotation, leading to an increase in thermal noise.
Proper tilt sensors and actuators should be placed for the purpose of feedback control.
The achievable RMS amplitude is eventually limited by the noise level of the tilt sensor, which can be about $\sim10^{-9}$ rad RMS for optical levers \cite{BRS}.
Assuming this noise level, cross-coupling transfer can be reduced down to around $5\times10^{-11}$ rad/m.
Interferometric sensors can be more sensitive to tilt, with potential sensitivities of $\sim10^{-12}$ rad, leading to around $5\times10^{-14}$ rad/m coupling level in principle.
The latter value meets the requirement for TOBA with a suspension point vibration of $10^{-6}$ m/$\sqrt{\rm Hz}$, which is the same as ground vibration and therefore can be realized without vibration isolation systems.

We also have to take care of other cross-coupling routes neglected in our measurements such as the Yaw tilt of the mirror surface $\phi_{\rm Y}$.
The cross-coupling transfer function of this route is $\frac{\phi_{\rm Y}}{L} ( \tilde{T}_{{\rm T}\rightarrow{\rm T}} - \tilde{B}_{{\rm T}\rightarrow{\rm T}} )$.
Assuming $\phi_{\rm Y} =10^{-6}$ rad and the transfer function to relative translation is $\tilde{T}_{{\rm T}\rightarrow{\rm T}} - \tilde{B}_{{\rm T}\rightarrow{\rm T}} =10^{-3}$ m/m at 0.1 Hz, the cross-coupling level is $5\times10^{-9}$ rad/m at 0.1 Hz.
A problem is that the relevant parameters about this route are difficult to improve significantly. 
Enhancing the surface quality of the mirror can reduce the coupling, but realistically the relative tilt $\phi_{\rm Y}$ will increase for longer bars.
Therefore the factor $\frac{\phi_{\rm Y}}{L}$ cannot be expected to decrease by much. 
A more effective way would be suppressing the relative translation between the bar and the optics, which is achieved by suspending the optics in the same way as the bar.
If the transfer functions $\tilde{T}_{{\rm T}\rightarrow{\rm T}}$ and $\tilde{B}_{{\rm T}\rightarrow{\rm T}}$ are completely identical, the coupling transfer will then be zero.
Unfortunately, small differences of realistic suspension parameters such as resonant frequencies will cause differential motion.
The current $10^{-3}$ m/m assumption is derived from a 5 \% difference of the translational resonant frequency.
As a realistic estimation, we can expect a 0.1 \% difference which is realized by adjusting the CoM within 0.5 mm accuracy for a 250 mm length suspension wire.
This asymmetry corresponds to a cross-coupling level of $1\times10^{-10}$ rad/m.
Further suppression will require careful design including consideration of the rotational resonant frequency and damping coefficients, but in general matching the all suspension parameters is quite difficult since the bar and the optical bench have different geometries.
Therefore, additional reduction to at least about $10^{-3}$ of the current level is required for the future TOBA sensitivity.
This can be achieved with active vibration isolation to the order of $10^{-9}$ m/$\sqrt{\rm Hz}$.
Another possible option is monitoring relative translation with an auxiliary sensor and then subtracting it from the Yaw signal.

\section{\label{sec:conc}Conclusion}
We have summarized how seismic cross-coupling transfers are introduced in a simple torsion pendulum.
The routes of coupling are common to both optical levers and Michelson interferometers, which are usually used in torsion pendulum experiments. 
Coupling transfer functions are also calculated for each route.
They are found to be approximately proportional to Pitch/Roll rotations or Trans/Vert translations, with proportional factors coming from the tilt of the system.
Therefore it is clear that cross-coupling can be reduced by adjusting the tilt of the system.

These calculations, along with the scheme to reduce the cross-coupling, were experimentally demonstrated with a two-stage torsion pendulum.
Coupling transfer functions were measured in two horizontal directions, Long and Trans, and they agreed with our calculations within experimental errors.
The tilt of the system was estimated by fitting the transfer functions.
Based on the estimated tilt, we placed counter weights on the TM and the OB, and also used actuators on the TM.
We succeeded in reducing coupling transfer functions in line with theory.
Thus we demonstrated not only the reduction of cross-coupling but also the validity of the theory.
Finally, coupling transfer levels of $5\times10^{-6}$ rad/m for Long and $4\times10^{-5}$ rad/m for Trans at 0.1 Hz were achieved.
This is an important milestone for geophysical applications since the strain sensitivity can reach to about $10^{-14} /\sqrt{\rm Hz}$, which is close to expected Newtonian noise and earthquake gravity signals, with these coupling levels and active vibration isolation.
Achievable coupling values can be improved by suppressing the RMS amplitude of Roll/Pitch tilt with a more sensitive tilt sensor.
We also note that for the final target sensitivity of TOBA, the route which was insignificant in our current experiment can dominate and hence vibration isolation to at least about $10^{-9}$ m/$\sqrt{\rm Hz}$ will be required.
This will be possible to achieve with an active vibration isolation system.
In conclusion, these results have established a basic way to reduce cross-coupling in torsion pendulums for the most significant coupling routes, and also elucidated on the prospects for seismic noise reduction of TOBA.

\section*{\label{sec:ack}Acknowledgement}
This work was supported by JSPS KAKENHI Grants Number JP16H03972, JP24244031 and JP18684005.
We thank Ooi Ching Pin for editing this article.


\bibliographystyle{apsrev4-1}

\begin{thebibliography}{27}%
\makeatletter
\providecommand \@ifxundefined [1]{%
 \@ifx{#1\undefined}
}%
\providecommand \@ifnum [1]{%
 \ifnum #1\expandafter \@firstoftwo
 \else \expandafter \@secondoftwo
 \fi
}%
\providecommand \@ifx [1]{%
 \ifx #1\expandafter \@firstoftwo
 \else \expandafter \@secondoftwo
 \fi
}%
\providecommand \natexlab [1]{#1}%
\providecommand \enquote  [1]{``#1''}%
\providecommand \bibnamefont  [1]{#1}%
\providecommand \bibfnamefont [1]{#1}%
\providecommand \citenamefont [1]{#1}%
\providecommand \href@noop [0]{\@secondoftwo}%
\providecommand \href [0]{\begingroup \@sanitize@url \@href}%
\providecommand \@href[1]{\@@startlink{#1}\@@href}%
\providecommand \@@href[1]{\endgroup#1\@@endlink}%
\providecommand \@sanitize@url [0]{\catcode `\\12\catcode `\$12\catcode
  `\&12\catcode `\#12\catcode `\^12\catcode `\_12\catcode `\%12\relax}%
\providecommand \@@startlink[1]{}%
\providecommand \@@endlink[0]{}%
\providecommand \url  [0]{\begingroup\@sanitize@url \@url }%
\providecommand \@url [1]{\endgroup\@href {#1}{\urlprefix }}%
\providecommand \urlprefix  [0]{URL }%
\providecommand \Eprint [0]{\href }%
\providecommand \doibase [0]{http://dx.doi.org/}%
\providecommand \selectlanguage [0]{\@gobble}%
\providecommand \bibinfo  [0]{\@secondoftwo}%
\providecommand \bibfield  [0]{\@secondoftwo}%
\providecommand \translation [1]{[#1]}%
\providecommand \BibitemOpen [0]{}%
\providecommand \bibitemStop [0]{}%
\providecommand \bibitemNoStop [0]{.\EOS\space}%
\providecommand \EOS [0]{\spacefactor3000\relax}%
\providecommand \BibitemShut  [1]{\csname bibitem#1\endcsname}%
\let\auto@bib@innerbib\@empty
\bibitem [{\citenamefont {{B. P. Abbott {\it et
  al.,}}}(2016{\natexlab{a}})}]{GW150914}%
  \BibitemOpen
  \bibfield  {author} {\bibinfo {author} {\bibnamefont {{B. P. Abbott {\it et
  al.,}}}} (\bibinfo {collaboration} {LIGO Scientific Collaboration and Virgo
  Collaboration}),\ }\href {\doibase 10.1103/PhysRevLett.116.061102} {\bibfield
   {journal} {\bibinfo  {journal} {Phys. Rev. Lett.}\ }\textbf {\bibinfo
  {volume} {116}},\ \bibinfo {pages} {061102} (\bibinfo {year}
  {2016}{\natexlab{a}})}\BibitemShut {NoStop}%
\bibitem [{\citenamefont {{B. P. Abbott {\it et
  al.,}}}(2016{\natexlab{b}})}]{GW151226}%
  \BibitemOpen
  \bibfield  {author} {\bibinfo {author} {\bibnamefont {{B. P. Abbott {\it et
  al.,}}}} (\bibinfo {collaboration} {LIGO Scientific Collaboration and Virgo
  Collaboration}),\ }\href {\doibase 10.1103/PhysRevLett.116.241103} {\bibfield
   {journal} {\bibinfo  {journal} {Phys. Rev. Lett.}\ }\textbf {\bibinfo
  {volume} {116}},\ \bibinfo {pages} {241103} (\bibinfo {year}
  {2016}{\natexlab{b}})}\BibitemShut {NoStop}%
\bibitem [{\citenamefont {{B. P. Abbott {\it et
  al.,}}}(2017{\natexlab{a}})}]{GW170104}%
  \BibitemOpen
  \bibfield  {author} {\bibinfo {author} {\bibnamefont {{B. P. Abbott {\it et
  al.,}}}} (\bibinfo {collaboration} {LIGO Scientific and Virgo
  Collaboration}),\ }\href {\doibase 10.1103/PhysRevLett.118.221101} {\bibfield
   {journal} {\bibinfo  {journal} {Phys. Rev. Lett.}\ }\textbf {\bibinfo
  {volume} {118}},\ \bibinfo {pages} {221101} (\bibinfo {year}
  {2017}{\natexlab{a}})}\BibitemShut {NoStop}%
\bibitem [{\citenamefont {{B. P. Abbott {\it et
  al.,}}}(2017{\natexlab{b}})}]{GW170814}%
  \BibitemOpen
  \bibfield  {author} {\bibinfo {author} {\bibnamefont {{B. P. Abbott {\it et
  al.,}}}} (\bibinfo {collaboration} {LIGO Scientific Collaboration and Virgo
  Collaboration}),\ }\href {\doibase 10.1103/PhysRevLett.119.141101} {\bibfield
   {journal} {\bibinfo  {journal} {Phys. Rev. Lett.}\ }\textbf {\bibinfo
  {volume} {119}},\ \bibinfo {pages} {141101} (\bibinfo {year}
  {2017}{\natexlab{b}})}\BibitemShut {NoStop}%
\bibitem [{\citenamefont {{B. P. Abbott {\it et
  al.,}}}(2017{\natexlab{c}})}]{GW170817}%
  \BibitemOpen
  \bibfield  {author} {\bibinfo {author} {\bibnamefont {{B. P. Abbott {\it et
  al.,}}}} (\bibinfo {collaboration} {LIGO Scientific Collaboration and Virgo
  Collaboration}),\ }\href {\doibase 10.1103/PhysRevLett.119.161101} {\bibfield
   {journal} {\bibinfo  {journal} {Phys. Rev. Lett.}\ }\textbf {\bibinfo
  {volume} {119}},\ \bibinfo {pages} {161101} (\bibinfo {year}
  {2017}{\natexlab{c}})}\BibitemShut {NoStop}%
\bibitem [{\citenamefont {{J Aasi {\it et al.,} (LIGO Scientific
  Collaboration)}}(2015)}]{LIGO}%
  \BibitemOpen
  \bibfield  {author} {\bibinfo {author} {\bibnamefont {{J Aasi {\it et al.,}
  (LIGO Scientific Collaboration)}}},\ }\href
  {http://stacks.iop.org/0264-9381/32/i=7/a=074001} {\bibfield  {journal}
  {\bibinfo  {journal} {Classical and Quantum Gravity}\ }\textbf {\bibinfo
  {volume} {32}},\ \bibinfo {pages} {074001} (\bibinfo {year}
  {2015})}\BibitemShut {NoStop}%
\bibitem [{\citenamefont {{F. Acernese {\it et al.,} (Virgo
  Collaboration)}}(2015)}]{Virgo}%
  \BibitemOpen
  \bibfield  {author} {\bibinfo {author} {\bibnamefont {{F. Acernese {\it et
  al.,} (Virgo Collaboration)}}},\ }\href
  {http://stacks.iop.org/0264-9381/32/i=2/a=024001} {\bibfield  {journal}
  {\bibinfo  {journal} {Classical and Quantum Gravity}\ }\textbf {\bibinfo
  {volume} {32}},\ \bibinfo {pages} {024001} (\bibinfo {year}
  {2015})}\BibitemShut {NoStop}%
\bibitem [{\citenamefont {{T Akutsu {\it et al.,} (KAGRA
  Collaboration)}}(2018)}]{iKAGRA}%
  \BibitemOpen
  \bibfield  {author} {\bibinfo {author} {\bibnamefont {{T Akutsu {\it et al.,}
  (KAGRA Collaboration)}}},\ }\href {\doibase 10.1093/ptep/ptx180} {\bibfield
  {journal} {\bibinfo  {journal} {Progress of Theoretical and Experimental
  Physics}\ }\textbf {\bibinfo {volume} {2018}},\ \bibinfo {pages} {013F01}
  (\bibinfo {year} {2018})}\BibitemShut {NoStop}%
\bibitem [{\citenamefont {Ando}\ \emph {et~al.}(2010)\citenamefont {Ando},
  \citenamefont {Ishidoshiro}, \citenamefont {Yamamoto}, \citenamefont {Yagi},
  \citenamefont {Kokuyama}, \citenamefont {Tsubono},\ and\ \citenamefont
  {Takamori}}]{TOBA}%
  \BibitemOpen
  \bibfield  {author} {\bibinfo {author} {\bibfnamefont {M.}~\bibnamefont
  {Ando}}, \bibinfo {author} {\bibfnamefont {K.}~\bibnamefont {Ishidoshiro}},
  \bibinfo {author} {\bibfnamefont {K.}~\bibnamefont {Yamamoto}}, \bibinfo
  {author} {\bibfnamefont {K.}~\bibnamefont {Yagi}}, \bibinfo {author}
  {\bibfnamefont {W.}~\bibnamefont {Kokuyama}}, \bibinfo {author}
  {\bibfnamefont {K.}~\bibnamefont {Tsubono}}, \ and\ \bibinfo {author}
  {\bibfnamefont {A.}~\bibnamefont {Takamori}},\ }\href {\doibase
  10.1103/PhysRevLett.105.161101} {\bibfield  {journal} {\bibinfo  {journal}
  {Phys. Rev. Lett.}\ }\textbf {\bibinfo {volume} {105}},\ \bibinfo {pages}
  {161101} (\bibinfo {year} {2010})}\BibitemShut {NoStop}%
\bibitem [{\citenamefont {{K. Danzmann and the LISA study team}}(1996)}]{LISA}%
  \BibitemOpen
  \bibfield  {author} {\bibinfo {author} {\bibnamefont {{K. Danzmann and the
  LISA study team}}},\ }\href {http://stacks.iop.org/0264-9381/13/i=11A/a=033}
  {\bibfield  {journal} {\bibinfo  {journal} {Classical and Quantum Gravity}\
  }\textbf {\bibinfo {volume} {13}},\ \bibinfo {pages} {A247} (\bibinfo {year}
  {1996})}\BibitemShut {NoStop}%
\bibitem [{\citenamefont {{Seiji Kawamura {\it et al.,}}}(2011)}]{DECIGO}%
  \BibitemOpen
  \bibfield  {author} {\bibinfo {author} {\bibnamefont {{Seiji Kawamura {\it et
  al.,}}}},\ }\href {http://stacks.iop.org/0264-9381/28/i=9/a=094011}
  {\bibfield  {journal} {\bibinfo  {journal} {Classical and Quantum Gravity}\
  }\textbf {\bibinfo {volume} {28}},\ \bibinfo {pages} {094011} (\bibinfo
  {year} {2011})}\BibitemShut {NoStop}%
\bibitem [{\citenamefont {Saulson}(1984)}]{NN}%
  \BibitemOpen
  \bibfield  {author} {\bibinfo {author} {\bibfnamefont {P.~R.}\ \bibnamefont
  {Saulson}},\ }\href {\doibase 10.1103/PhysRevD.30.732} {\bibfield  {journal}
  {\bibinfo  {journal} {Phys. Rev. D}\ }\textbf {\bibinfo {volume} {30}},\
  \bibinfo {pages} {732} (\bibinfo {year} {1984})}\BibitemShut {NoStop}%
\bibitem [{\citenamefont {Hughes}\ and\ \citenamefont {Thorne}(1998)}]{NNseis}%
  \BibitemOpen
  \bibfield  {author} {\bibinfo {author} {\bibfnamefont {S.~A.}\ \bibnamefont
  {Hughes}}\ and\ \bibinfo {author} {\bibfnamefont {K.~S.}\ \bibnamefont
  {Thorne}},\ }\href {\doibase 10.1103/PhysRevD.58.122002} {\bibfield
  {journal} {\bibinfo  {journal} {Phys. Rev. D}\ }\textbf {\bibinfo {volume}
  {58}},\ \bibinfo {pages} {122002} (\bibinfo {year} {1998})}\BibitemShut
  {NoStop}%
\bibitem [{\citenamefont {Fiorucci}\ \emph {et~al.}(2018)\citenamefont
  {Fiorucci}, \citenamefont {Harms}, \citenamefont {Barsuglia}, \citenamefont
  {Fiori},\ and\ \citenamefont {Paoletti}}]{NNatomD}%
  \BibitemOpen
  \bibfield  {author} {\bibinfo {author} {\bibfnamefont {D.}~\bibnamefont
  {Fiorucci}}, \bibinfo {author} {\bibfnamefont {J.}~\bibnamefont {Harms}},
  \bibinfo {author} {\bibfnamefont {M.}~\bibnamefont {Barsuglia}}, \bibinfo
  {author} {\bibfnamefont {I.}~\bibnamefont {Fiori}}, \ and\ \bibinfo {author}
  {\bibfnamefont {F.}~\bibnamefont {Paoletti}},\ }\href {\doibase
  10.1103/PhysRevD.97.062003} {\bibfield  {journal} {\bibinfo  {journal} {Phys.
  Rev. D}\ }\textbf {\bibinfo {volume} {97}},\ \bibinfo {pages} {062003}
  (\bibinfo {year} {2018})}\BibitemShut {NoStop}%
\bibitem [{\citenamefont {Driggers}\ \emph {et~al.}(2012)\citenamefont
  {Driggers}, \citenamefont {Harms},\ and\ \citenamefont
  {Adhikari}}]{NNcancel}%
  \BibitemOpen
  \bibfield  {author} {\bibinfo {author} {\bibfnamefont {J.~C.}\ \bibnamefont
  {Driggers}}, \bibinfo {author} {\bibfnamefont {J.}~\bibnamefont {Harms}}, \
  and\ \bibinfo {author} {\bibfnamefont {R.~X.}\ \bibnamefont {Adhikari}},\
  }\href {\doibase 10.1103/PhysRevD.86.102001} {\bibfield  {journal} {\bibinfo
  {journal} {Phys. Rev. D}\ }\textbf {\bibinfo {volume} {86}},\ \bibinfo
  {pages} {102001} (\bibinfo {year} {2012})}\BibitemShut {NoStop}%
\bibitem [{\citenamefont {Harms}\ \emph {et~al.}(2015)\citenamefont {Harms},
  \citenamefont {Ampuero}, \citenamefont {Barsuglia}, \citenamefont
  {Chassande-Mottin}, \citenamefont {Montagner}, \citenamefont {Somala},\ and\
  \citenamefont {Whiting}}]{Earthquake}%
  \BibitemOpen
  \bibfield  {author} {\bibinfo {author} {\bibfnamefont {J.}~\bibnamefont
  {Harms}}, \bibinfo {author} {\bibfnamefont {J.-P.}\ \bibnamefont {Ampuero}},
  \bibinfo {author} {\bibfnamefont {M.}~\bibnamefont {Barsuglia}}, \bibinfo
  {author} {\bibfnamefont {E.}~\bibnamefont {Chassande-Mottin}}, \bibinfo
  {author} {\bibfnamefont {J.-P.}\ \bibnamefont {Montagner}}, \bibinfo {author}
  {\bibfnamefont {S.~N.}\ \bibnamefont {Somala}}, \ and\ \bibinfo {author}
  {\bibfnamefont {B.~F.}\ \bibnamefont {Whiting}},\ }\href {\doibase
  10.1093/gji/ggv090} {\bibfield  {journal} {\bibinfo  {journal} {Geophysical
  Journal International}\ }\textbf {\bibinfo {volume} {201}},\ \bibinfo {pages}
  {1416} (\bibinfo {year} {2015})}\BibitemShut {NoStop}%
\bibitem [{\citenamefont {Montagner}\ \emph {et~al.}(2016)\citenamefont
  {Montagner}, \citenamefont {Juhel}, \citenamefont {Barsuglia}, \citenamefont
  {Ampuero}, \citenamefont {Chassande-Mottin}, \citenamefont {Harms},
  \citenamefont {Whiting}, \citenamefont {Bernard}, \citenamefont
  {Cl{\'e}v{\'e}d{\'e}},\ and\ \citenamefont {Lognonn{\'e}}}]{EEW_JPM2016}%
  \BibitemOpen
  \bibfield  {author} {\bibinfo {author} {\bibfnamefont {J.-P.}\ \bibnamefont
  {Montagner}}, \bibinfo {author} {\bibfnamefont {K.}~\bibnamefont {Juhel}},
  \bibinfo {author} {\bibfnamefont {M.}~\bibnamefont {Barsuglia}}, \bibinfo
  {author} {\bibfnamefont {J.~P.}\ \bibnamefont {Ampuero}}, \bibinfo {author}
  {\bibfnamefont {E.}~\bibnamefont {Chassande-Mottin}}, \bibinfo {author}
  {\bibfnamefont {J.}~\bibnamefont {Harms}}, \bibinfo {author} {\bibfnamefont
  {B.}~\bibnamefont {Whiting}}, \bibinfo {author} {\bibfnamefont
  {P.}~\bibnamefont {Bernard}}, \bibinfo {author} {\bibfnamefont
  {E.}~\bibnamefont {Cl{\'e}v{\'e}d{\'e}}}, \ and\ \bibinfo {author}
  {\bibfnamefont {P.}~\bibnamefont {Lognonn{\'e}}},\ }\href
  {http://dx.doi.org/10.1038/ncomms13349} {\bibfield  {journal} {\bibinfo
  {journal} {Nature Communications}\ }\textbf {\bibinfo {volume} {7}},\
  \bibinfo {pages} {13349 EP } (\bibinfo {year} {2016})}\BibitemShut {NoStop}%
\bibitem [{\citenamefont {Vall{\'e}e}\ \emph {et~al.}(2017)\citenamefont
  {Vall{\'e}e}, \citenamefont {Ampuero}, \citenamefont {Juhel}, \citenamefont
  {Bernard}, \citenamefont {Montagner},\ and\ \citenamefont
  {Barsuglia}}]{EEW_MV2017}%
  \BibitemOpen
  \bibfield  {author} {\bibinfo {author} {\bibfnamefont {M.}~\bibnamefont
  {Vall{\'e}e}}, \bibinfo {author} {\bibfnamefont {J.~P.}\ \bibnamefont
  {Ampuero}}, \bibinfo {author} {\bibfnamefont {K.}~\bibnamefont {Juhel}},
  \bibinfo {author} {\bibfnamefont {P.}~\bibnamefont {Bernard}}, \bibinfo
  {author} {\bibfnamefont {J.-P.}\ \bibnamefont {Montagner}}, \ and\ \bibinfo
  {author} {\bibfnamefont {M.}~\bibnamefont {Barsuglia}},\ }\href {\doibase
  10.1126/science.aao0746} {\bibfield  {journal} {\bibinfo  {journal}
  {Science}\ }\textbf {\bibinfo {volume} {358}},\ \bibinfo {pages} {1164}
  (\bibinfo {year} {2017})},\ \Eprint
  {http://arxiv.org/abs/http://science.sciencemag.org/content/358/6367/1164.full.pdf}
  {http://science.sciencemag.org/content/358/6367/1164.full.pdf} \BibitemShut
  {NoStop}%
\bibitem [{\citenamefont {Ishidoshiro}\ \emph {et~al.}(2011)\citenamefont
  {Ishidoshiro}, \citenamefont {Ando}, \citenamefont {Takamori}, \citenamefont
  {Takahashi}, \citenamefont {Okada}, \citenamefont {Matsumoto}, \citenamefont
  {Kokuyama}, \citenamefont {Kanda}, \citenamefont {Aso},\ and\ \citenamefont
  {Tsubono}}]{Phase1TOBA}%
  \BibitemOpen
  \bibfield  {author} {\bibinfo {author} {\bibfnamefont {K.}~\bibnamefont
  {Ishidoshiro}}, \bibinfo {author} {\bibfnamefont {M.}~\bibnamefont {Ando}},
  \bibinfo {author} {\bibfnamefont {A.}~\bibnamefont {Takamori}}, \bibinfo
  {author} {\bibfnamefont {H.}~\bibnamefont {Takahashi}}, \bibinfo {author}
  {\bibfnamefont {K.}~\bibnamefont {Okada}}, \bibinfo {author} {\bibfnamefont
  {N.}~\bibnamefont {Matsumoto}}, \bibinfo {author} {\bibfnamefont
  {W.}~\bibnamefont {Kokuyama}}, \bibinfo {author} {\bibfnamefont
  {N.}~\bibnamefont {Kanda}}, \bibinfo {author} {\bibfnamefont
  {Y.}~\bibnamefont {Aso}}, \ and\ \bibinfo {author} {\bibfnamefont
  {K.}~\bibnamefont {Tsubono}},\ }\href {\doibase
  10.1103/PhysRevLett.106.161101} {\bibfield  {journal} {\bibinfo  {journal}
  {Phys. Rev. Lett.}\ }\textbf {\bibinfo {volume} {106}},\ \bibinfo {pages}
  {161101} (\bibinfo {year} {2011})}\BibitemShut {NoStop}%
\bibitem [{\citenamefont {Shoda}\ \emph {et~al.}(2017)\citenamefont {Shoda},
  \citenamefont {Kuwahara}, \citenamefont {Ando}, \citenamefont {Eda},
  \citenamefont {Tejima}, \citenamefont {Aso},\ and\ \citenamefont
  {Itoh}}]{Phase2TOBA}%
  \BibitemOpen
  \bibfield  {author} {\bibinfo {author} {\bibfnamefont {A.}~\bibnamefont
  {Shoda}}, \bibinfo {author} {\bibfnamefont {Y.}~\bibnamefont {Kuwahara}},
  \bibinfo {author} {\bibfnamefont {M.}~\bibnamefont {Ando}}, \bibinfo {author}
  {\bibfnamefont {K.}~\bibnamefont {Eda}}, \bibinfo {author} {\bibfnamefont
  {K.}~\bibnamefont {Tejima}}, \bibinfo {author} {\bibfnamefont
  {Y.}~\bibnamefont {Aso}}, \ and\ \bibinfo {author} {\bibfnamefont
  {Y.}~\bibnamefont {Itoh}},\ }\href {\doibase 10.1103/PhysRevD.95.082004}
  {\bibfield  {journal} {\bibinfo  {journal} {Phys. Rev. D}\ }\textbf {\bibinfo
  {volume} {95}},\ \bibinfo {pages} {082004} (\bibinfo {year}
  {2017})}\BibitemShut {NoStop}%
\bibitem [{\citenamefont {Collette}\ and\ \citenamefont
  {Matichard}(2015)}]{AVI}%
  \BibitemOpen
  \bibfield  {author} {\bibinfo {author} {\bibfnamefont {C.}~\bibnamefont
  {Collette}}\ and\ \bibinfo {author} {\bibfnamefont {F.}~\bibnamefont
  {Matichard}},\ }\href {\doibase https://doi.org/10.1016/j.jsv.2015.01.006}
  {\bibfield  {journal} {\bibinfo  {journal} {Journal of Sound and Vibration}\
  }\textbf {\bibinfo {volume} {342}},\ \bibinfo {pages} {1 } (\bibinfo {year}
  {2015})}\BibitemShut {NoStop}%
\bibitem [{\citenamefont {Harms}\ and\ \citenamefont
  {Mow-Lowry}(2018)}]{sp-asp}%
  \BibitemOpen
  \bibfield  {author} {\bibinfo {author} {\bibfnamefont {J.}~\bibnamefont
  {Harms}}\ and\ \bibinfo {author} {\bibfnamefont {C.~M.}\ \bibnamefont
  {Mow-Lowry}},\ }\href {http://stacks.iop.org/0264-9381/35/i=2/a=025008}
  {\bibfield  {journal} {\bibinfo  {journal} {Classical and Quantum Gravity}\
  }\textbf {\bibinfo {volume} {35}},\ \bibinfo {pages} {025008} (\bibinfo
  {year} {2018})}\BibitemShut {NoStop}%
\bibitem [{\citenamefont {McManus}\ \emph {et~al.}(2017)\citenamefont
  {McManus}, \citenamefont {Forsyth}, \citenamefont {Yap}, \citenamefont
  {Ward}, \citenamefont {Shaddock}, \citenamefont {McClelland},\ and\
  \citenamefont {Slagmolen}}]{TorPeDO}%
  \BibitemOpen
  \bibfield  {author} {\bibinfo {author} {\bibfnamefont {D.~J.}\ \bibnamefont
  {McManus}}, \bibinfo {author} {\bibfnamefont {P.~W.~F.}\ \bibnamefont
  {Forsyth}}, \bibinfo {author} {\bibfnamefont {M.~J.}\ \bibnamefont {Yap}},
  \bibinfo {author} {\bibfnamefont {R.~L.}\ \bibnamefont {Ward}}, \bibinfo
  {author} {\bibfnamefont {D.~A.}\ \bibnamefont {Shaddock}}, \bibinfo {author}
  {\bibfnamefont {D.~E.}\ \bibnamefont {McClelland}}, \ and\ \bibinfo {author}
  {\bibfnamefont {B.~J.~J.}\ \bibnamefont {Slagmolen}},\ }\href
  {http://stacks.iop.org/0264-9381/34/i=13/a=135002} {\bibfield  {journal}
  {\bibinfo  {journal} {Classical and Quantum Gravity}\ }\textbf {\bibinfo
  {volume} {34}},\ \bibinfo {pages} {135002} (\bibinfo {year}
  {2017})}\BibitemShut {NoStop}%
\bibitem [{\citenamefont {Hoyle}\ \emph {et~al.}(2004)\citenamefont {Hoyle},
  \citenamefont {Kapner}, \citenamefont {Heckel}, \citenamefont {Adelberger},
  \citenamefont {Gundlach}, \citenamefont {Schmidt},\ and\ \citenamefont
  {Swanson}}]{Washington2}%
  \BibitemOpen
  \bibfield  {author} {\bibinfo {author} {\bibfnamefont {C.~D.}\ \bibnamefont
  {Hoyle}}, \bibinfo {author} {\bibfnamefont {D.~J.}\ \bibnamefont {Kapner}},
  \bibinfo {author} {\bibfnamefont {B.~R.}\ \bibnamefont {Heckel}}, \bibinfo
  {author} {\bibfnamefont {E.~G.}\ \bibnamefont {Adelberger}}, \bibinfo
  {author} {\bibfnamefont {J.~H.}\ \bibnamefont {Gundlach}}, \bibinfo {author}
  {\bibfnamefont {U.}~\bibnamefont {Schmidt}}, \ and\ \bibinfo {author}
  {\bibfnamefont {H.~E.}\ \bibnamefont {Swanson}},\ }\href {\doibase
  10.1103/PhysRevD.70.042004} {\bibfield  {journal} {\bibinfo  {journal} {Phys.
  Rev. D}\ }\textbf {\bibinfo {volume} {70}},\ \bibinfo {pages} {042004}
  (\bibinfo {year} {2004})}\BibitemShut {NoStop}%
\bibitem [{\citenamefont {Harms}\ and\ \citenamefont
  {Venkateswara}(2016)}]{NNcancel_tiltmeter}%
  \BibitemOpen
  \bibfield  {author} {\bibinfo {author} {\bibfnamefont {J.}~\bibnamefont
  {Harms}}\ and\ \bibinfo {author} {\bibfnamefont {K.}~\bibnamefont
  {Venkateswara}},\ }\href {http://stacks.iop.org/0264-9381/33/i=23/a=234001}
  {\bibfield  {journal} {\bibinfo  {journal} {Classical and Quantum Gravity}\
  }\textbf {\bibinfo {volume} {33}},\ \bibinfo {pages} {234001} (\bibinfo
  {year} {2016})}\BibitemShut {NoStop}%
\bibitem [{\citenamefont {Arellano}\ \emph {et~al.}(2013)\citenamefont
  {Arellano}, \citenamefont {Panjwani}, \citenamefont {Carbone},\ and\
  \citenamefont {Speake}}]{AngularSensor_FEPA}%
  \BibitemOpen
  \bibfield  {author} {\bibinfo {author} {\bibfnamefont {F.~E.~P.}\
  \bibnamefont {Arellano}}, \bibinfo {author} {\bibfnamefont {H.}~\bibnamefont
  {Panjwani}}, \bibinfo {author} {\bibfnamefont {L.}~\bibnamefont {Carbone}}, \
  and\ \bibinfo {author} {\bibfnamefont {C.~C.}\ \bibnamefont {Speake}},\
  }\href {\doibase 10.1063/1.4795549} {\bibfield  {journal} {\bibinfo
  {journal} {Review of Scientific Instruments}\ }\textbf {\bibinfo {volume}
  {84}},\ \bibinfo {pages} {043101} (\bibinfo {year} {2013})},\ \Eprint
  {http://arxiv.org/abs/https://doi.org/10.1063/1.4795549}
  {https://doi.org/10.1063/1.4795549} \BibitemShut {NoStop}%
\bibitem [{\citenamefont {Venkateswara}\ \emph {et~al.}(2014)\citenamefont
  {Venkateswara}, \citenamefont {Hagedorn}, \citenamefont {Turner},
  \citenamefont {Arp},\ and\ \citenamefont {Gundlach}}]{BRS}%
  \BibitemOpen
  \bibfield  {author} {\bibinfo {author} {\bibfnamefont {K.}~\bibnamefont
  {Venkateswara}}, \bibinfo {author} {\bibfnamefont {C.~A.}\ \bibnamefont
  {Hagedorn}}, \bibinfo {author} {\bibfnamefont {M.~D.}\ \bibnamefont
  {Turner}}, \bibinfo {author} {\bibfnamefont {T.}~\bibnamefont {Arp}}, \ and\
  \bibinfo {author} {\bibfnamefont {J.~H.}\ \bibnamefont {Gundlach}},\ }\href
  {\doibase 10.1063/1.4862816} {\bibfield  {journal} {\bibinfo  {journal}
  {Review of Scientific Instruments}\ }\textbf {\bibinfo {volume} {85}},\
  \bibinfo {pages} {015005} (\bibinfo {year} {2014})}\BibitemShut {NoStop}%
\end{thebibliography}
%

\end{document}